\documentclass[format=acmlarge,review=false]{acmart}

\usepackage{booktabs} 
\usepackage{pdfcomment}
\definecolor{red}{rgb}{0,0,0}
\usepackage{subcaption}

\usepackage{array}
\makeatletter
\providecommand{\tabularnewline}{\\}
\makeatother

\usepackage[ruled]{algorithm2e} 

\SetAlFnt{\small}
\SetAlCapFnt{\small}
\SetAlCapNameFnt{\small}
\SetAlCapHSkip{0pt}
\IncMargin{-\parindent}

\acmConference{IMWUT}
\acmVolume{3}
\acmNumber{1}
\acmArticle{0}
\acmYear{2019}
\acmMonth{6}
\acmArticleSeq{0}

\setcopyright{usgovmixed}

\acmDOI{0000001.0000001}

\received{November 2018}
\received{February 2019}
\received[accepted]{April 2019}

\begin{document}

\title{PocketCare: Tracking the Flu with Mobile Phones using Partial Observations of Proximity and Symptoms}

\author{Wen Dong}
\orcid{0000-0001-8923-2227}
\email{wendong@buffalo.edu}
\affiliation{University at Buffalo}
\author{Tong Guan}
\email{tongguan@buffalo.edu}
\affiliation{University at Buffalo}
\author{Bruno Lepri}
\email{lepri@fbk.eu}
\affiliation{Fondazione Bruno Kessler}
\author{Chunming Qiao}
\email{qiao@buffalo.edu}
\affiliation{University at Buffalo}

\begin{abstract}
Mobile phones provide a powerful sensing platform that researchers may adopt to understand proximity interactions among people and the diffusion, through these interactions, of diseases, behaviors, and opinions. However, it remains a challenge to track the proximity-based interactions of a whole community and then model the social diffusion of diseases and behaviors starting from the observations of a small fraction of the volunteer population. 
In this paper, we propose a novel approach that tries to connect together these sparse observations using a model of how individuals interact with each other and how social interactions happen in terms of a sequence of proximity interactions. We apply our approach to track the spreading of flu in the spatial-proximity network of a 3000-people university campus by mobilizing 300 volunteers from this population to monitor nearby mobile phones through Bluetooth scanning and to daily report flu symptoms about and around them. Our aim is to predict the likelihood for an individual to get flu based on how often her/his daily routine intersects with those of the volunteers. Thus, we use the daily routines of the volunteers to build a model of the volunteers as well as of the non-volunteers. Our results show that we can predict flu infection two weeks ahead of time with an average precision from 0.24 to 0.35 depending on the amount of information. This precision is six to nine times higher than with a  {\color{red}random guess} model. At the population level, we can predict infectious population in a two-week window with an r-squared value of 0.95 (a {\color{red}random-guess} model obtains an r-squared value of 0.2). These results point to an innovative approach for tracking individuals who have interacted with people showing symptoms, allowing us to warn those in danger of infection and to inform health researchers about the progression of contact-induced diseases.
\end{abstract}

%
%

\ccsdesc[500]{Human-centered computing~Empirical studies in collaborative and social computing}
\ccsdesc[500]{Human-centered computing~Ubiquitous and mobile computing systems and tools}
\ccsdesc[500]{Human-centered computing~Empirical studies in ubiquitous and mobile computing}
\ccsdesc[500]{Computing methodologies~Machine learning}
\ccsdesc[500]{Computing methodologies~Discrete-event simulation}
%
%

\keywords{Social diffusion, social network dynamics, sensor network,  signal processing, machine learning}

\maketitle

\section{Introduction}
Mobile phones provide a powerful sensing platform that researchers in social science, epidemiology, social psychology, marketing, etc. may adopt to understand proximity interactions among people and the diffusion, through these interactions, of diseases \cite{madan2010social,dong2012modeling,dong2012graph,farrahi2014epidemic}, eating habits \cite{madan2010eating}, opinions \cite{madan2010eating}, behaviors \cite{pan2011composite}, emotions, etc.
However, it remains a challenge to track the proximity-based interactions of a whole community or organization and model the social diffusion of behaviors, diseases, and opinions starting from the observations of a small fraction of the volunteer population, and it is extremely hard to provide the right incentives to the whole organization and community for participating to the data collection. Thus, being unable to observe or infer the individual-level interactions at large-scale imposes serious limitations in the living lab-based studies because the solution of many societal problems involves the observation of not only the volunteers but also the non-volunteers. For example, being able to control epidemics involves knowing diseases spreading among the observed as well as the unobserved individuals in the social network.

Our proposed solution tries to connect together the sparse data points, obtained from volunteers' mobile phones, using a model of how individuals interact with each other and how social interactions happen in terms of a sequence of elementary events (e.g., a sequence of proximity interactions). Each of these elementary events involves only few individuals but in sequence the events let emerge complex behaviors. More specifically, we proceed inferring the latent state of the complex system and calibrating the probabilities of the elementary events as a function of the latent state, using the sparse data points contributed by the volunteers as observations of the latent process. In this way, we combine techniques from generative machine learning, signal processing, and agent-based modeling to reveal the dynamics of the whole organization or community.  

In the current paper, we describe the application of this approach to track the spreading of flu in the spatial-proximity network of a 3000-people university campus by mobilizing 300 volunteers from this population to monitor nearby mobile phones through Bluetooth scanning every 5 minutes and to daily report flu symptoms about and around them. Our aim is to predict the likelihood for an individual to get flu based on how often her/his daily routine intersects with those of the volunteers. Specifically, while the non-volunteers are not observable, their routines will share some similar patterns with the ones of the volunteers from the same community, and their spatial proximity with one another is detected through the Bluetooth scanning of the volunteers' mobile phones. Thus, we use the daily routines of the volunteers to build a model of the volunteers as well as of the non-volunteers among the population: In particular, we model when they stay at the apartments; how they move from one class to the next one; when they take their meals and go to the gym; how many people spend time in proximity of apartments, classes, food court and gym, and so on. Then, we use the observations of the volunteers' behaviors and proximity interactions as constraints for the model to unroll over time: When, where and how large the classes are for individuals to mix, how common cold and flu spread among the individuals to give rise to the daily symptoms reported by the volunteers, and how the unobserved part of the proximity network is wired to fit the reported symptoms and the observed routines.   

Our work contributes to the research of predicting the spatial-proximity interactions among individuals in a community from the observations of a small set of volunteers. More specifically, we derive machine learning algorithms for a discrete-event model to infer the latent interactions from isolated observations. The premise of introducing the discrete-event model is that many complex social-interaction dynamics can be factored into a sequence of elementary events which individually involve only a few individuals but together bring the complex behavior of the system (e.g., the dyadic interactions and infections monitored by the volunteers). We provide evidence that this approach of stitching together isolated observations by volunteers with high-fidelity social interaction models may build an affordable ``microscope" for computational social scientists and policy researchers to study our social systems.

The remainder of this paper is organized as follows. In Section 2, we discuss other research on capturing proximity interactions through mobile phones and wearable devices, and on modeling the spreading of behaviors, diseases, opinions, and psychological states. In Section 3 we introduce PocketCare, the Android sensing platform we developed to track proximity interactions, locations, and behaviors, and describe the data sets collected. In Section 4, we present our discrete-event model, the so-called stochastic kinetic model used to capture the spatial-proximity network dynamics, and the inference and learning algorithms. Section 5 reports the results obtained in tracking the spread of the flu, while in Section 6 we discuss the implications and limitations of our work and draw some conclusions.

\section{Literature Review}

Over the past decade, mobile phones \cite{eagle2006reality,eagle2009inferring,aharony2011social,laurila2012mobile,stopczynski2014measuring,wang2014studentlife,centellegher2016mobile} and other wearable devices \cite{olguin2009sensible,lepri2012sociometric,alshamsi2015beyond} have been increasingly used as tools for closely observing individual and community behaviors. In one of the earliest studies, Eagle and Pentland analyzed the routines \cite{eagle2006reality} and the social interactions \cite{eagle2009inferring} of various research teams at the MIT Media Laboratory and the MIT Sloan School of Management by means of 84 Nokia 6600 smart phones. Similarly, using the multimodal data (i.e., infra-red, accelerometer, Bluetooth, and audio) captured by sociometric badges \cite{olguin2009sensible}, Olguin \emph{et al.} studied the information flow, efficiency, and employee satisfaction of various organizations. Aharony \emph{et al.} observed the daily lives of 130 adult members of a young-family residential living community for 15 months using the FunF Android app \cite{aharony2011social}. In Nokia data-collection challenge, 185 participants shared five months of data about their daily activities and social interactions for scientific research \cite{laurila2012mobile}. More recently, Stopczynski \emph{et al.} \cite{stopczynski2014measuring} conducted the Copenhagen Networks Study on a densely connected population of 1000 individuals for multiple years, using mobile phones to collect data on face-to-face and communication interactions, social networks, locations, and background information such as demographics and personality traits. Table \ref{tab:studies} summarizes the previous research on capturing spatial-proximity interactions and the adopted mobile sensing and wearable platforms. Our approach stands out for monitoring the social interactions of a significantly larger community of individuals by mobilizing only a small number of volunteers. 

\begin{table}
\caption{Research efforts for monitoring spatial-proximity interactions}~\label{tab:studies}

\begin{tabular}{|>{\centering}p{0.135\textwidth}|>{\centering}p{0.2\textwidth}|>{\centering}p{0.19\textwidth}|>{\centering}p{0.09\textwidth}|>{\centering}p{0.27\textwidth}|}
\hline 
Research Study & Subjects & Observed Behavior & Platform & Signals\tabularnewline
\hline 
\hline 
Reality Mining \cite{eagle2006reality} & Loosely coupled teams of 2-10 people each & Routines and social interactions & Nokia 6600 & Sensor and survey data about participants for 9 months\tabularnewline
\hline 
Sociometric Badges \cite{olguin2009sensible,lederman2017open} & Teams of 10-20 people & Work-space dynamics & Proprietary hardware & Sensor and survey data about participants for 1-14 days\tabularnewline
\hline 
Social Patterns \cite{genois2015compensating} & Communities of 100 people & Spatial-proximity interactions & Proprietary hardware & Sensor and survey data about participants for 1-7 days\tabularnewline
\hline 
Social Evolution \cite{madan2010social} & 84 undergraduates in a student dormitory & Network dynamics and social diffusion & Android & Sensor and survey data about participants for 9 months\tabularnewline
\hline 
Friends \& Family \cite{aharony2011social} & 130 adult members of a dormitory & Influence and intervention & Android & Sensor and survey data about participants for 12 months\tabularnewline
\hline 
Nokia Data Collection \cite{laurila2012mobile} & 185 participants & Mobile crowd sensing & Android & Sensor and survey data about participants for two years\tabularnewline
\hline 
Copenhagen Networks \cite{stopczynski2014measuring} & 1000 incoming undergraduates & Spatial-proximity interactions & Android & Sensor and survey data about participants for multiple years\tabularnewline
\hline 
AWARE \cite{ferreira2015aware,van2016measuring} & 15 students in one study & Spatial-proximity interactions & Android, iOS & Sensor and survey data about participants for an hour\tabularnewline
\hline 
Student Life \cite{lane2011bewell,wang2014studentlife} & 48 students in a single class & Students' well-being and performance & Android & Sensor and survey data about participants for 5 months\tabularnewline
\hline 
PocketCare & 3000 people in a university department & Spatial-proximity interactions & Android & Sensor and survey data from 300 volunteers for multiple
years\tabularnewline
\hline 
\end{tabular}

\end{table}

Modeling and predicting the dissemination of opinions and diseases in a community is emerging as an important reason to capture face-to-face, proximity, and communication interactions by means of mobile phones and wearable devices. In particular, three main approaches have been proposed to deal with social dissemination: 1) a data-driven approach for predicting diffusion from a set of features, 2) a simulation approach for specifying high-fidelity diffusion dynamics based on Monte Carlo methods, and 3) an analytical approach for specifying diffusion dynamics by means of a tractable generative model and inferring the latent diffusion process from the observations. 

Among the data-driven approaches, Pan \emph{et al.} demonstrated the influence of social interactions on mobile-phone app installation by fitting multimodal social network data with an exponential random graph model \cite{pan2011composite}. Madan \emph{et al.} showed that long-time exposure to obese and inactive people influences gaining weight \cite{madan2010social}. Lane \emph{et al.} found that the prediction of behaviors, habits, and psychological states --- such as transportation choice, hours of sleep, diet, and mood --- are improved by also considering the behaviors, habits, and psychological states of the social network's neighbors \cite{lane2014connecting}. Along this line, Sandstrom \emph{et al.} found that emotional well-being is dependent on not only friends and family but also weaker social ties \cite{sandstrom2014social}. Finally, the results of a study by Wu \emph{et al.} \cite{wu2008mining} offered evidence that employee productivity in a workspace is dependent on the productivity of the face-to-face network's neighbors. However, it is worth noting that data-driven approaches require the collection of a sufficient amount of training data regarding the state of a network's neighbors in order to identify predictive features and train a classifier. Thus, these approaches become difficult to apply when the training data is scarce --- for example, in non-recurrent situations. 

Simulation modeling is widely used in policy research, where insights and optimal policies are identified by running computer simulation and Monte Carlo integration. State-of-the-art simulators of social diffusion include EpiSims \cite{eubank2004modelling} and STEM \cite{edlund2010spatiotemporal}. However, a big problem with simulating social diffusion through high-resolution spatial-proximity information collected with wearable sensors is that the network is often incomplete. G{\'e}nois \emph{et al.} demonstrated a method to alleviate the issue of using incomplete data about human interaction networks in social diffusion simulations by resampling the network \cite{genois2015compensating}. Fournet and Barrat proposed another method to augment an incomplete network by resampling new edges using surveyed friendship information \cite{fournet2014}. Farrahi \emph{et al.} \cite{farrahi2014epidemic} developed an approach using phone communication as a proxy for spatial-proximity in epidemic tracing. Barrat \emph{et al.} showed that the spreading pathways of the epidemic process are strongly affected by the temporal structure of the network data \cite{barrat2013empirical}. Finally, Lee \emph{et al.} demonstrated that immunization protocol based on a temporal contact network outperforms a static network, and therefore that the temporal contact structure has more information to exploit for developing a vaccination protocol \cite{lee2012exploiting}. 

Another approach for modeling dissemination through proximity interaction is to define a generative model of infection at the individual level as a hidden Markov process and infer the latent infection process from the observations of self-reported symptoms. With this approach, Dong \emph{et al.} developed a graph-coupled hidden Markov model to predict the spread of infection from one individual to another using the susceptible-infectious-susceptible (SIS) dynamics and the dynamics of the spatial-proximity network captured by mobile-phone Bluetooth scanning \cite{dong2012graph,dong2012modeling}. They adopted this approach to model the spatial-proximity network dynamics of more than 80\% of residents in an undergraduate student dormitory, where the residents spent more than 10 hours interacting with other people in the community. They dealt with the incompleteness of the observed dynamic proximity network by introducing an event to represent an infection from outside the network. Fan \emph{et al.} then developed a variational inference algorithm for graph-coupled hidden Markov models to make faster infection predictions \cite{fan2016unifying}. More recently, Xu \emph{et al.} introduced the stochastic kinetic model to generalize the graph-coupled hidden Markov model and derived a variational inference algorithm to accelerate infection prediction \cite{xu2016using}. 

A common element of these previous works is that they require a high-quality spatial-proximity network to be able to predict the spreading of an infection from one node to another in the network. A useful process is to infer both the social diffusion and the network dynamics from incomplete information, such that diffusion can be tracked among not only the observed volunteers but also the non-volunteers.

In the remainder of this paper, we describe our efforts to mobilize a small fraction of a population to monitor the presence of other mobile phones (and their owners) through Bluetooth scanning and signs of social diffusion (symptoms of the common cold and flu), and use machine learning algorithms and generative models to stitch together these isolated observations.

\begin{figure}[]
\centering
  \includegraphics[width=0.24\columnwidth]{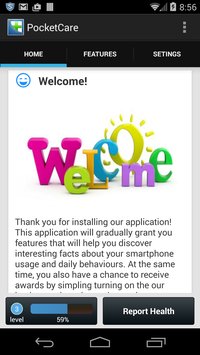}
  \includegraphics[width=0.24\columnwidth]{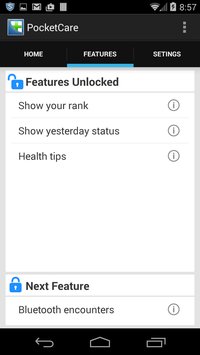}
  \includegraphics[width=0.24\columnwidth]{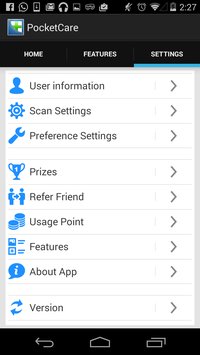}
  \includegraphics[width=0.24\columnwidth]{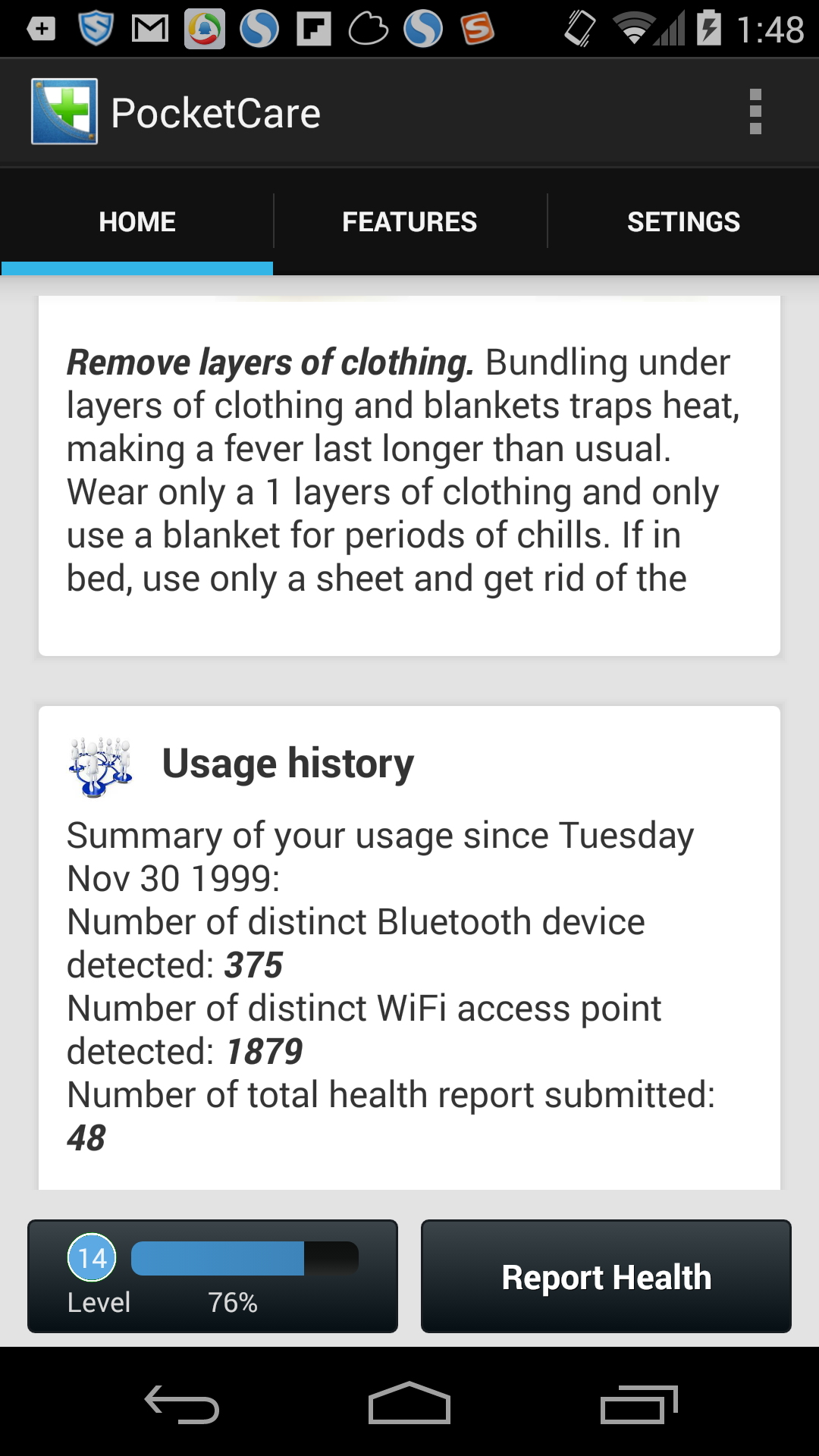}
  \caption{PocketCare is an app that monitors flu propagation and helps users to improve health by regularly tracking proximity network data and asking users to voluntarily report their flu symptoms and symptom observations about nearby people. }~\label{fig:pocketcare}
\end{figure}

\section{Mobile Sensing Platform and Collected Data}

We developed the Android app \emph{PocketCare} (see screenshots in Figure \ref{fig:pocketcare}) to monitor flu propagation and help users track and improve their health. To this end, PocketCare continuously 1) tracks a user's proximity network by periodically scanning for other Bluetooth devices within a few meters, including other mobile phones running PocketCare, 2) records the approximate location from GPS or nearby WiFi access points, depending on the user's settings, and 3) determines the user's activities (i.e., stationary, walking, or in a moving vehicle). In addition, PocketCare asks users to voluntarily report if they personally have flu symptoms and or have observed flu symptoms in people nearby, and provides users with useful health tips.

As an incentive for the study participants to provide information on flu symptoms, they accumulate reward points by giving the app permission to collect data. These points unlock additional useful features as well as prizes. The app runs in the background with negligible impact on normal phone usage or battery consumption.

Data are collected anonymously and do not reveal any personal information about the user. The collected data are stored on a secure server that can be accessed only by authorized researchers (although study participants also have the option to store such data locally on the phone). PocketCare adheres strictly to the privacy policies reviewed and approved by the university's Social and Behavioral Sciences IRB (SBSIRB), which can be found online.

Using PocketCare, we have now monitored flu propagation and provided flu information on a university campus for over a year. Approximately 300 users continuously operate PocketCare on campus: 108 graduate students, 144 undergraduate students, and 24 faculty/staff members from one university department with a total population of around 3000, as well as 64 users not from the department but from the same university. We selected 80\% of participants from the same department to capture a sufficient number of interactions and establish the ground truth. 
We chose the other 20\% of participants from the wider university population to evaluate the potential for tracking epidemics at larger scale from incomplete information about both social networks and symptoms using a mixture of generative machine learning models and agent-based models. 

During the study, we ask participants daily whether they personally are experiencing local symptoms (such as sore throat, sneezing, runny nose, or cough) or systemic symptoms (such as headache, fever, or muscle pains), and whether they have observed others sneezing or coughing. Different symptoms give us meaningful information about cold and flu development and progression, as the cold and flu syndromes typically involve early symptoms of headache, sneezing, and chilliness that are distinct from later symptoms such as coughing \cite{jackson1958transmission}. The survey is a conventional tool used by epidemiologists --- well-studied and widely applied in discovering epidemic progression at the population level.

\begin{figure}[]
\centering
  \includegraphics[width=0.45\columnwidth]{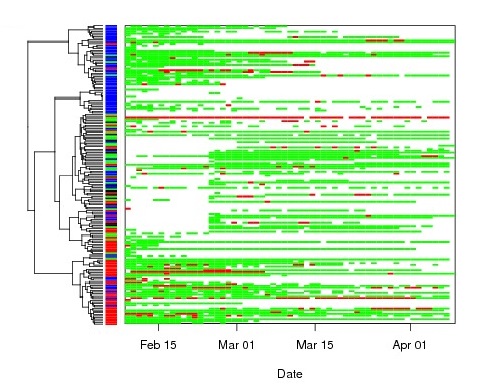}
  \includegraphics[width=0.45\columnwidth]{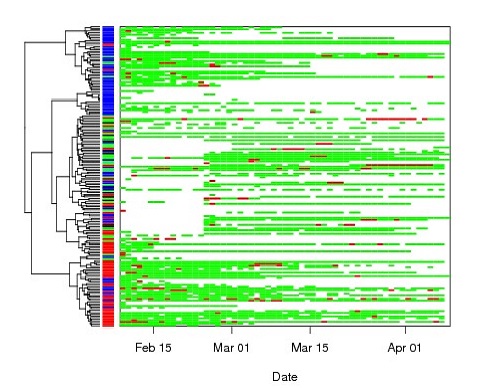}
  \caption{Reported symptoms of PocketCare users (left) and symptoms of nearby people observed by users (right) show the flu spreading through a spatial-proximity network. Rows are indexed by study participants, and columns by dates. Red, green, and blank cells respectively represent the existence of symptoms, the absence of symptoms, and unknown circumstances. The rows are permuted according to a hierarchical clustering algorithm using the Ward agglomeration method and the cosine distance to bring spatial-proximity interactions closer.}~\label{fig:sick}\vspace{-.2in} 

\end{figure}

About 70\% of study participants fill out symptom surveys, and approximately 4\% of surveyed reports of symptoms say either that the participants have symptoms or that they have observed symptoms in others (Figure \ref{fig:sick}). When participants observe symptoms nearby, they are twice as likely to have symptoms themselves. We collected a sufficient number of symptom cases in our data set because people have on average two episodes of cold or flu per year, each lasting around one week, and these numbers are higher in younger people. Moreover, the symptoms recorded by the surveys are also the vector for spreading cold and flu, since they facilitate transmission \cite{monto2000clinical}. A cold or flu that causes a sub-clinical infection is unlikely to succeed in transmission, and the most successful common cold viruses are generally those that cause the most runny noses, coughs, and sneezes \cite{eccles2005asymptomatic}.

We use the Bluetooth interface with PocketCare to monitor nearby mobile phones, including those running PocketCare, and to establish a dynamic spatial-proximity network of mobile phones. The idea behind tracking spatial proximity is for one device to transmit signals through its Bluetooth transmitter while another measures the received signal strength indicator (RSSI) through its receiver. If the RSSI value is above a certain threshold, the app will consider the two mobile phone users to be in spatial proximity. After some initial calibration, the app can also estimate the distance between the two phones based on this RSSI value. 

\begin{figure*}[]
  \centering
  \begin{subfigure}[b]{0.33\textwidth}
      \centering
      \includegraphics[width=1\textwidth]{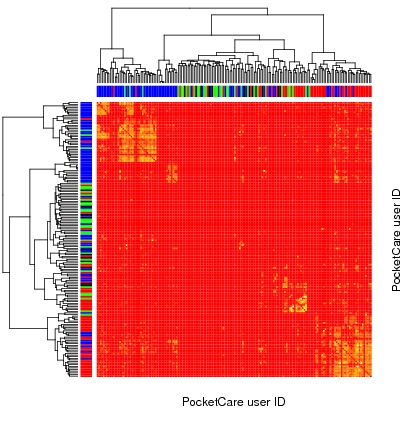}
      \caption{\label{fig:proximity-users} Bluetooth proximity of PocketCare users}
  \end{subfigure}
  \begin{subfigure}[b]{0.33\textwidth}
      \centering
      \includegraphics[trim=0 0 0 3cm, width=1\textwidth]{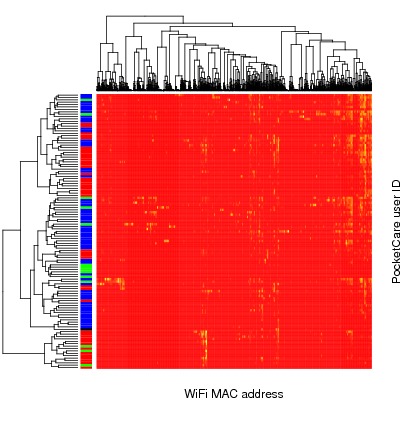}
      \caption{\label{fig:proximity-wlan}WiFi proximity of PocketCare users}
  \end{subfigure}
  \begin{subfigure}[b]{0.33\textwidth}
      \centering
      \includegraphics[width=1\textwidth]{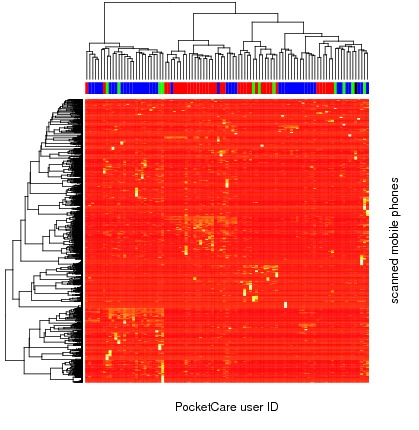}
      \caption{\label{fig:proximity-others}Bluetooth proximity of scanned phones}
  \end{subfigure}
  \caption{(\subref{fig:proximity-users}) Aggregated adjacency matrix of PocketCare users through Bluetooth scanning of nearby users (red - graduate, blue - undergraduate, black - staff, green - other). (\subref{fig:proximity-wlan}) The number of times PocketCare users saw on-campus WiFi access points. (\subref{fig:proximity-others}) Aggregated adjacency matrix of all phones in the same department Bluetooth-scanned by PocketCare users. Rows and columns are permuted according to a hierarchical clustering algorithm using the Ward agglomeration method and cosine distance.}~\label{fig:proximity}\vspace{-.2in}
\end{figure*}

PocketCare scans nearby Bluetooth devices every five minutes, and collects the anonymized PocketCare user ID, the Unix time when the scan was made, the media access control (MAC) addresses of nearby devices, and the RSSI values. The Bluetooth scanning records show three macro clusters among PocketCare users, corresponding to the graduates, the undergraduates, and the faculty/staff of the department, with each macro cluster consisting of smaller clusters according to where people live, which classes they take, which labs they belong to, and so on (Figure \ref{fig:proximity-users}). PocketCare-scanned mobile devices account for approximately 40\% of the department's population (Figure \ref{fig:proximity-others}) and about 15\% of the total population on campus, where we have used Bluetooth class of device/service and the organizationally unique identifier (OUI) in the MAC address to infer whether a device is a mobile phone, and the location of the PocketCare scanner to infer the location of scanned devices. Thus, the 300 PocketCare users monitor a spatial-proximity network far bigger than the network of the participants themselves. This is a sufficient level of coverage of a community to validate our modeling and machine learning algorithms.  

Additionally, we have used the WiFi interface with PocketCare to learn the structure of users' daily activities and their interactions in these activities. Scanning nearby WiFi access points every five minutes, PocketCare collects the anonymized PocketCare user ID, the Unix time when a scan was made, the MAC addresses of the WiFi access points, and the RSSI values. When combined with the geographic locations of the scanned WiFi access points (e.g., from the Wireless Geographic Logging Engine or \url{https://wigle.net/}) and information about those locations (e.g., from OpenStreetMap), the device scanning reveals a meaningful structure that can be used to synthesize agent activities and interactions for studying the spread of epidemics at the individual level. We can also train a classifier to identify from these WiFi scanning records whether two PocketCare users are in spatial proximity. For example, two users are likely to be in proximity if their RSSIs to the same access points are close and they have often seen the same access points at the same times in the past. Given the locations of the WiFi access points (e.g., their latitudes and longitudes, or their relative positions on a floor plan), we can use as few as nine WiFi scan records to calibrate the signal-decay models and compute users' locations with an average of two-meter accuracy.  

\begin{figure*}[p!]
  \centering
  \begin{subfigure}[b]{1\textwidth}
      \centering
      \includegraphics[width=.15\textwidth]{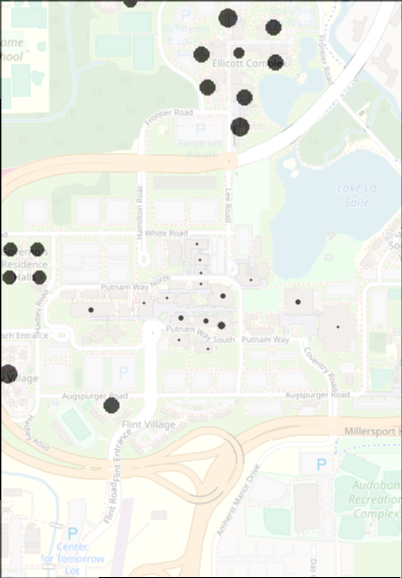}
      \includegraphics[width=.15\textwidth]{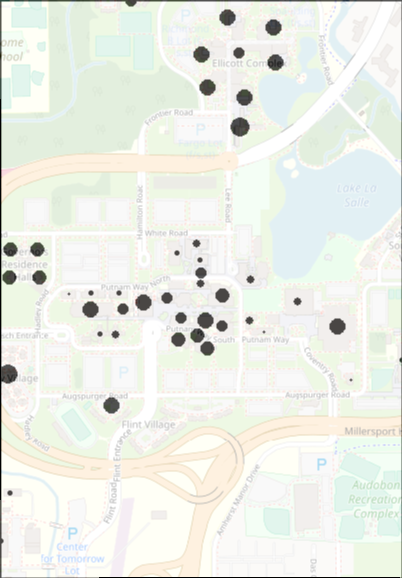}
      \includegraphics[width=.15\textwidth]{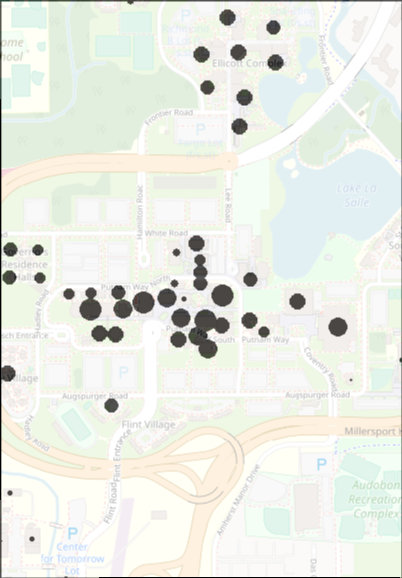}
      \includegraphics[width=.15\textwidth]{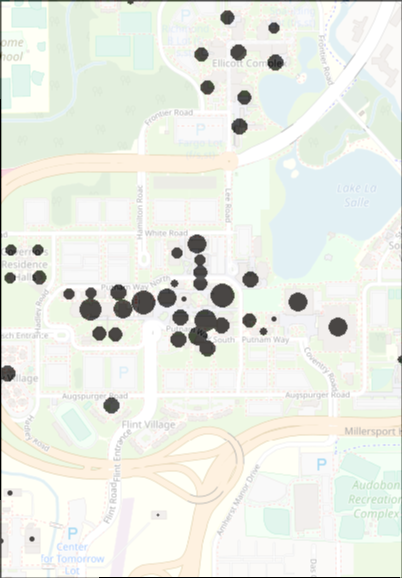}
      \includegraphics[width=.15\textwidth]{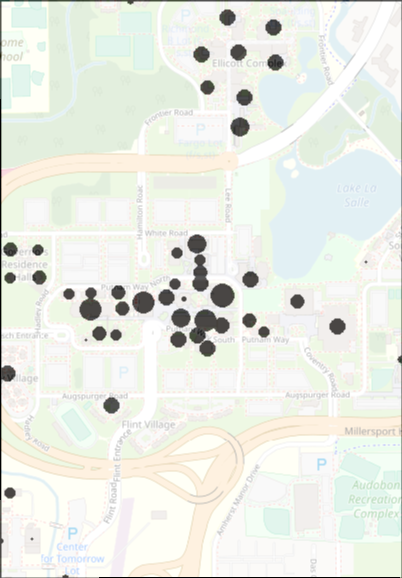}
      \includegraphics[width=.15\textwidth]{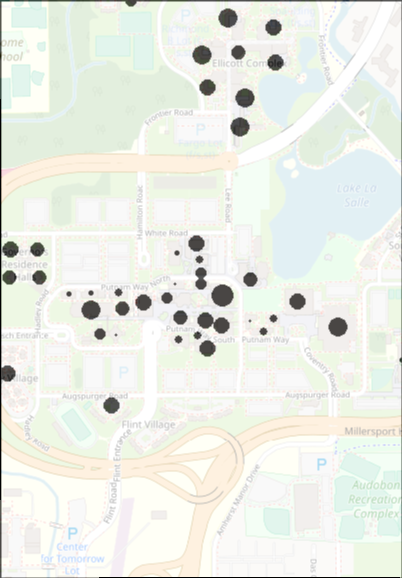}
      \caption{\label{fig:cit-weekday} Aggregated weekday activities of the entire university community across the campus (midnight, 8am, 10am, noon, 3pm, and 7pm).}
  \end{subfigure}
  \begin{subfigure}[b]{1\textwidth}
      \centering
      \includegraphics[width=.15\textwidth]{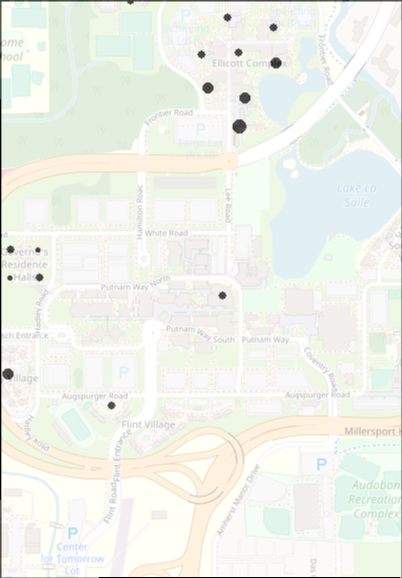}
      \includegraphics[width=.15\textwidth]{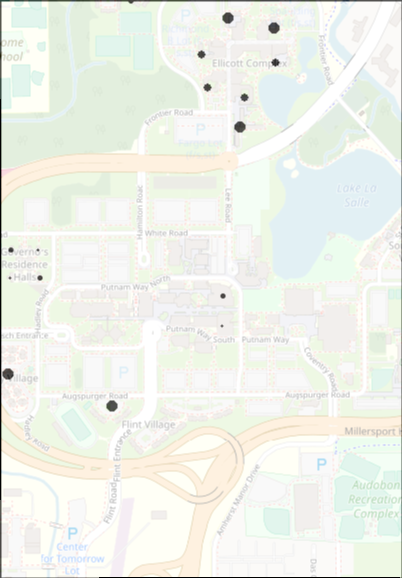}
      \includegraphics[width=.15\textwidth]{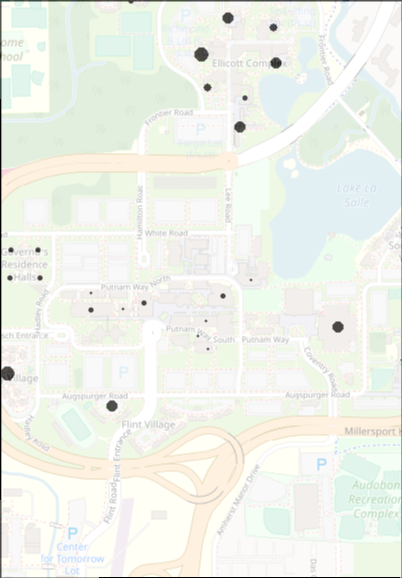}
      \includegraphics[width=.15\textwidth]{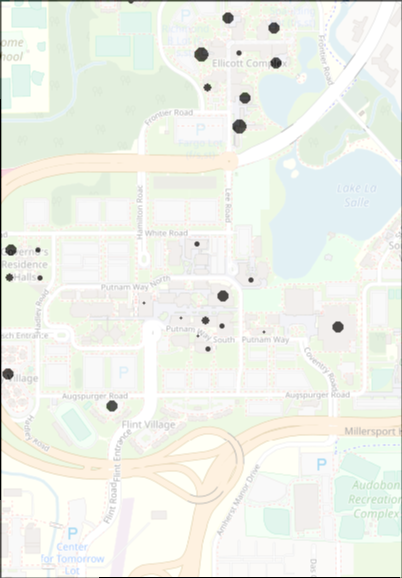}
      \includegraphics[width=.15\textwidth]{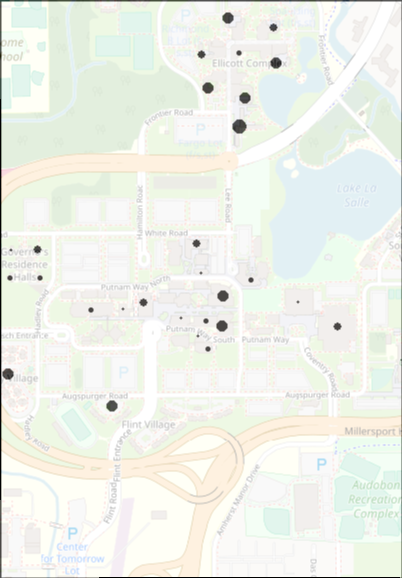}
      \includegraphics[width=.15\textwidth]{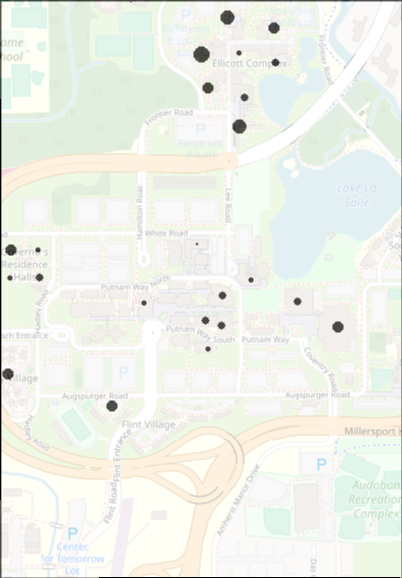}
      \caption{\label{fig:cit-weekend} Aggregated weekend activities of the entire university community across the campus (midnight, 8am, 10am, noon, 3pm, and 7pm).}
  \end{subfigure}
  \begin{subfigure}[b]{1\textwidth}
      \centering
      \includegraphics[width=.15\textwidth]{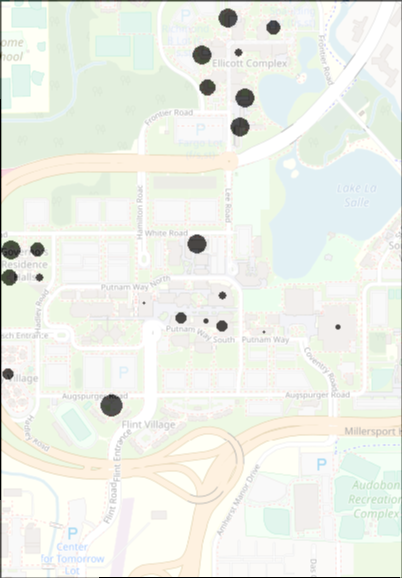}
      \includegraphics[width=.15\textwidth]{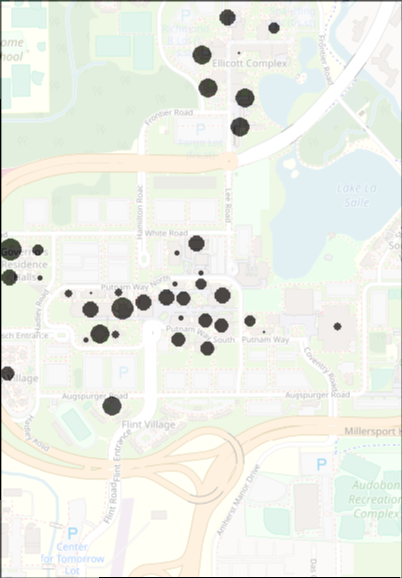}
      \includegraphics[width=.15\textwidth]{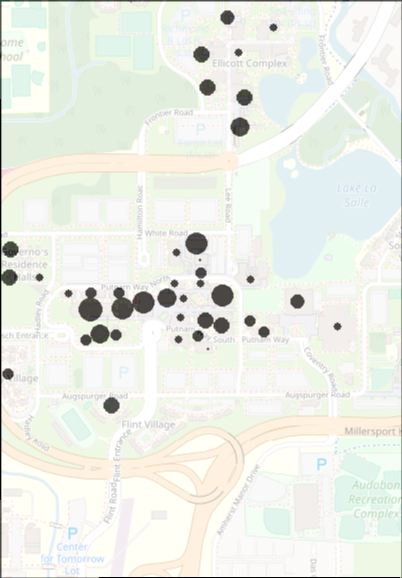}
      \includegraphics[width=.15\textwidth]{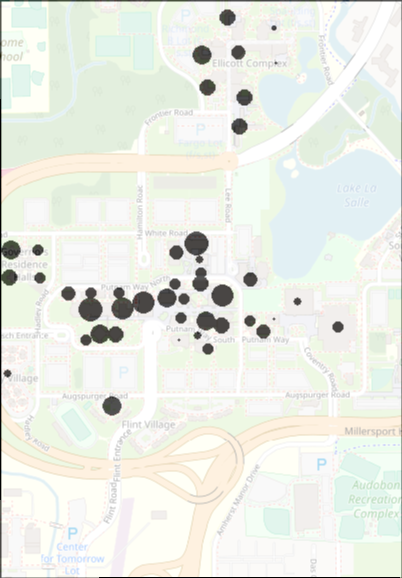}
      \includegraphics[width=.15\textwidth]{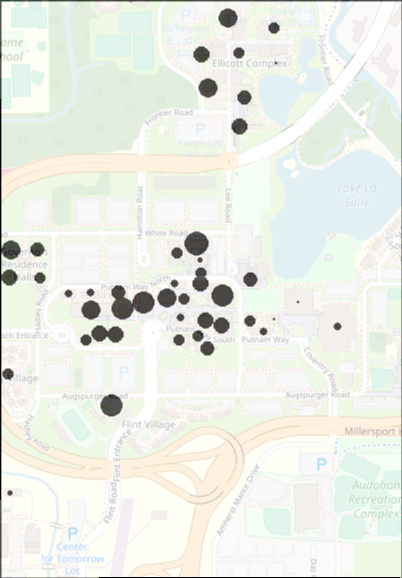}
      \includegraphics[width=.15\textwidth]{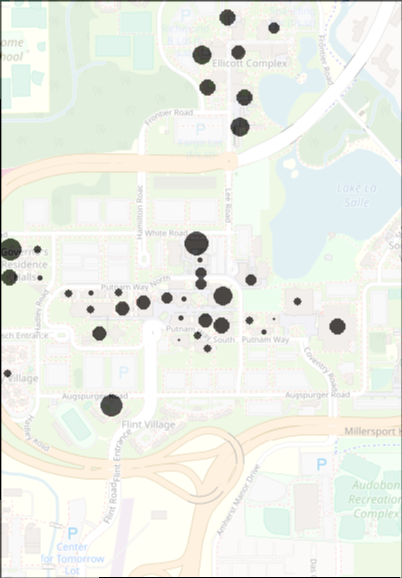}
      \caption{\label{fig:wifi-weekday} Aggregated weekday activities of PocketCare users across the campus (midnight, 8am, 10am, noon, 3pm and, 7pm).}
  \end{subfigure}
  \begin{subfigure}[b]{1\textwidth}
      \centering
      \includegraphics[width=.15\textwidth]{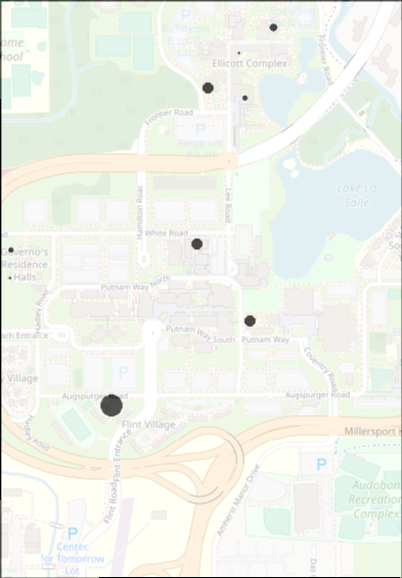}
      \includegraphics[width=.15\textwidth]{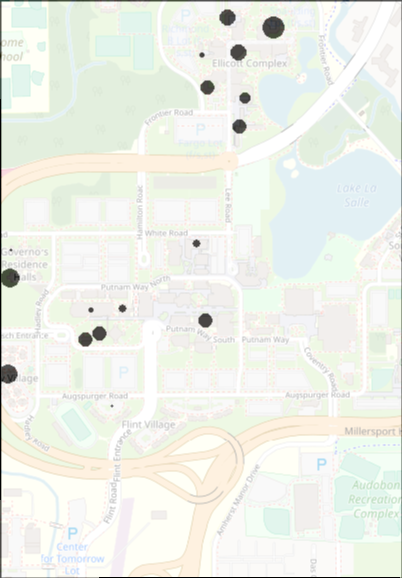}
      \includegraphics[width=.15\textwidth]{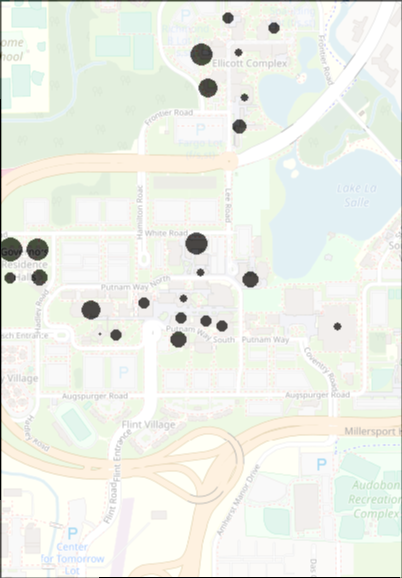}
      \includegraphics[width=.15\textwidth]{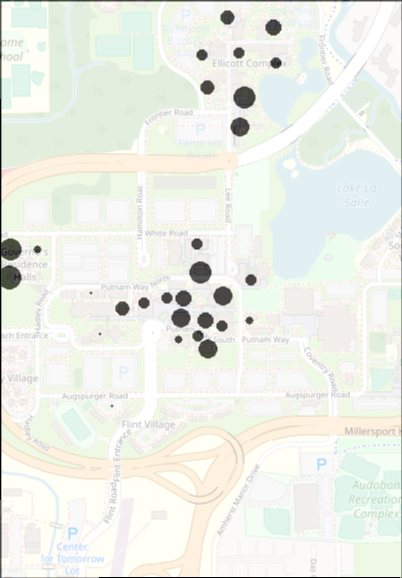}
      \includegraphics[width=.15\textwidth]{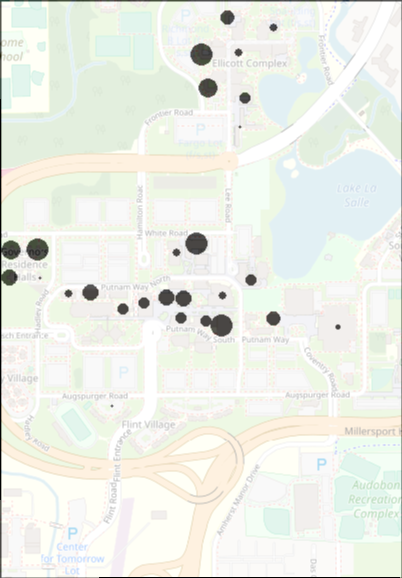}
      \includegraphics[width=.15\textwidth]{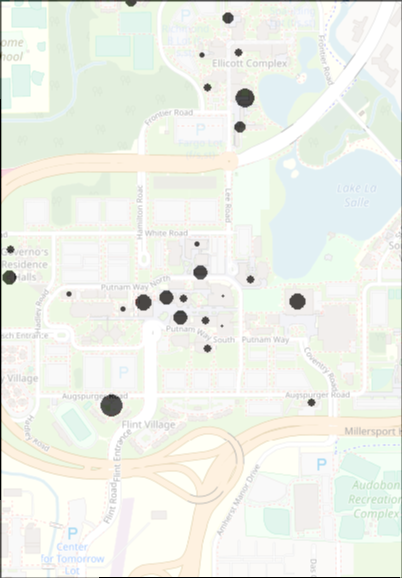}
      \caption{\label{fig:wifi-weekend} Aggregated weekend activities of PocketCare users across the campus (midnight, 8am, 10am, noon, 3pm, and 7pm).}
  \end{subfigure}
  \caption{The activities of the entire university community and of PocketCare users specifically on a typical weekday and weekend in terms of where their mobile phones are connected to WiFi access points. Dots of varying sizes in logarithmic scale represent populations in different on-campus buildings.}~\label{fig:wifi}
\end{figure*}

The WiFi access-point scanning records show three macro clusters and subclusters of PocketCare users, according to how long the users share spaces with one another (Figure \ref{fig:proximity-wlan}). In the data set, users on average frequent the same four to eight places per day, including labs, lecture and residence buildings, university libraries, and the student center, at similar hours of the day. However, some users can frequent as many as 40 places at the 90th percentile. Some places, such as student dormitories and office buildings, see the same set of PocketCare users, while others such as libraries and the student center see almost all users. This behavior agrees with previous findings made at a university dormitory \cite{dong2011modeling,guan2017fine}, and provides the basis for us to specify a generative model of the dynamic spatial-proximity network through how people move around and perform different activities throughout the day (Figures \ref{fig:wifi-weekday} and \ref{fig:wifi-weekend}).

To establish the ground truth of the campus-wide spatial-proximity network, we used syslog records containing which users are connected to which university WiFi access points at what times. These data contain the anonymized user ID, the anonymized device MAC address connecting to an access point, the OUI of the connecting device MAC address (which has information about device class), the MAC address of the connected access point, and the starting and ending times of the connection. This data set contains the interactions of about 60,000 devices from 30,000 users with 3,000 WiFi access points distributed to approximately 500 building floors. An average user frequents the same four to eight places per day at similar hours of a day, but a few users can frequent as many as 140 places on a specific day. A floor is generally frequented by the same two people to several hundreds of people every day, but some floors (for example, at the student center or libraries) can see different thousands of people daily. These behaviors are compatible with those in the PocketCare WiFi-scan data (Figures \ref{fig:cit-weekday} and \ref{fig:cit-weekend}).

Finally, we also monitored user activities every five minutes with the ActivityRecognitionApi Google API for Android in PocketCare. The device is still 86\% of the time, in use for 5\%, on a user who is walking or running for 3\%, and in a vehicle for 2\%. Thus, for a significant fraction of time users are likely indoors, and their locations and interactions are better captured by WiFi and Bluetooth than using GPS. 

While Bluetooth and Wi-Fi scanning are supposedly energy-intensive, battery-life issues caused by scanning have not been reported by the volunteers. The battery drain appears to be negligible.

\section{Methodology}
In this section, we introduce a discrete event model called the stochastic kinetic model to capture the dynamics of a complex social system, then we formulate social diffusion and spatial-proximity network dynamics with a discrete event model, and finally we offer inference and learning algorithms.

\subsection{Modeling}
We introduce the stochastic kinetic model to capture the dynamics of a complex social system driven by a set of events. A \emph{stochastic kinetic model} is a biochemist's way of describing the temporal evolution of a biological network with $M$ species driven by $V$ mutually independent events \cite{gillespie2007stochastic,wilkinson2011stochastic}, where the stochastic effects are particularly prevalent (e.g., a transcription network or a signal transduction network). Let $\mathbf X=(\mathbf X^{(1)},\cdots,\mathbf X^{(M)})$ denote individuals belonging to the $M$ species in the network. An event (chemical reaction) $v$ is specified by a production
\begin{align} 
 & {\alpha_{v}^{(1)}\mathbf X^{(1)}+\cdots+\alpha_{v}^{(M)}\mathbf X^{(M)}\overset{c_v}\to\beta_{v}^{(1)}\mathbf X^{(1)}+\cdots+\beta_{v}^{(M)}\mathbf X^{(M)}}.\label{eq:reaction}
\end{align}
The production is interpreted as having \emph{rate constant} $c_{v}$ (probability per unit time, as time goes to 0), $\alpha_{v}^{(1)}$ individuals of species $1$, $\alpha_{v}^{(2)}$ individuals of species $2$ ... interact according to event $v$, thus resulting in their removal from the system; and $\beta_{v}^{(1)}$ individuals of species $1$, $\beta_{v}^{(2)}$ individuals of species $2$ ... are introduced into the system. Hence, event $v$ changes the populations by $\Delta_{v}=(\beta_{v}^{(1)}- \alpha_{v}^{(1)},\cdots,\beta_{v}^{(M)}-\alpha_{v}^{(M)})$. The species on the left side of the production are \emph{reactants}, the species on the right side of the production are \emph{products}, and the species $m$ with $\alpha_v^{(m)}=\beta_v^{(m)}$ are \emph{catalysts}.

At the system level, let $x_{t}=(x_{t}^{(1)},\dots,x_{t}^{(M)})$ be the populations of the species in the system at time $t$. A stochastic kinetic process initially in state $x_{0}$ at time $t=0$ can be simulated through the Gillespie algorithm \cite{gillespie1976} (Algorithm \ref{alg:Gillespie}). In this algorithm, event rate $h_{v}(x_t,c_{v})$ is the rate constant $c_{v}$ multiplying a total of $\prod_{m=1}^{M}(x^{(m)}_t)^{\alpha_{v}^{(m)}}$ different ways for individuals to interact in the system, assuming homogeneous populations; and event rate takes a more complex form if the populations are not homogeneous. Exponential distribution is the maximum entropy distribution given the rate constant, and consequently is most likely to occur in natural reactions \cite{gillespie2007stochastic}. The stochastic kinetic model thus assigns a probabilistic measure to a sample path induced by a sequence of events $v_{1},\dots,v_{n}$ happening between times $0$ and $T$, $0<t_{1}<\dots<t_{n}<T$, which is 
\begin{align}
 & P(v_{1:n},t_{1:n},x_{1:n}) = \prod\limits _{i=1}^{n}h_{v_{i}}(x_{t_{i-1}},c_{v_{i}})\exp(-\sum\limits _{i=1}^{n}h_{0}(x_{t_{i-1}},c)(t_{i}-t_{i-1})),\label{eq:CTSKM} \\
  & \text{where }h_{v}(x,c_{k})=c_{v}g_{v}(x)\text{ for }v=1,\cdots,V,\text{ and  }h_{0}(x,c)=\sum\limits_{v=1}^{V}h_{v}(x,c_{v}). \label{eq:rate}
\end{align}

\begin{algorithm}[tb]

\textbf{Input}: A stochastic kinetic process specified by a set of independent events (Eq. \ref{eq:reaction} for $v=1,\cdots,V$) with initial populations $x_{t}=(x_{t}^{(1)},\dots,x_{t}^{(M)})$ at time $t=0$, and event rate $h_v(x_t,c_v)$ for each event $v$. Termination condition which is often simulating upper time $T$.
\vspace*{.5em}

\textbf{Output}: A sequence of events $v_1,\cdots,v_n$, the times of these events $t_0=0, t_1,\cdots,t_n$ as well as the corresponding populations $x_0, x_1,\cdots,x_n$.
\vspace*{.5em}

\textbf{Procedure}: Iterate the following steps until the termination condition is satisfied. 
\begin{enumerate}
\item Sample the time $\tau$ to the next event according to exponential distribution $\tau\sim\mbox{Exponential}(h_{0}(x_t,c))$, where $h_{0}(x,c)=\sum_{v=1}^{V}h_{v}(x_t,c_{v})$ is the rate of all events.
\item Sample the event $v$ according to categorical distribution $v\sim(\frac{h_{1}}{h_{0}},\dots,\frac{h_{V}}{h_{0}})$ conditional on event time $\tau$.
\item Update the system time $t\leftarrow t+\tau$ and populations $x\leftarrow x+\Delta_{v}$.
\end{enumerate}
\caption{\label{alg:Gillespie}Gillespie algorithm to simulate a stochastic kinetic process}
\end{algorithm}

The stochastic kinetic model is one way to define a discrete event process, and its equivalents in other fields include the stochastic Petri net \cite{marsan1994modelling,goss1998quantitative}, the system dynamics model \cite{forrester1961w}, the multi-agent model specified through a flow chart or state chart \cite{borshchev2013big}, and the production rule system \cite{newell1972human}. These equivalent models have the same power in capturing the dynamics of a system, but are different in their representations. As such, the stochastic kinetic model can also be used in the fields where there are equivalent models.

\subsection{Dynamics of social diffusion and the proximity network}
In this subsection, we formulate the dynamics of epidemic spread and the activity-based spatial-proximity network, building on previous research.

The common cold is the most common ubiquitous infectious disease, generally occurring in the winter and spring. Average adults get 2~3two or three colds per year, and average children 6-8six to eight. The common cold is caused by infection of the throat, sinuses, and larynx by one of over 200 virus strains through close contact with infected people or indirectly through contact with objects in the environment, followed by transfer to the mouth or nose. Symptoms include coughing, sore throat, runny nose, sneezing, headache, and fever, which can appear within two days after of exposure to the virus and last for seven to ten days \cite{centers2016common}. Risk factors include going to daycare, not sleeping well, and psychological stress \cite{allan2014prevention}.

We use SIS dynamics to capture the dynamic spreading of cold and flu through spatial proximity. In SIS dynamics, each individual is either infectious (I) or susceptible (S), and the system has three events: 1) an infectious individual in the network infects a susceptible individual in the network and turns that person infectious with rate constant $c_1^{(d)}$ (probability per unit time), 2) an infectious individual recovers and becomes susceptible again with rate constant $c_2^{(d)}$, and 3) a susceptible individual becomes infectious by contacting an infectious individual from outside the system with rate constant $c_3^{(d)}$. 

\vspace{-.1in}\begin{align*}
& I +S \to 2\times I, \text{infection, rate constant } = c_1^{(d)},\\
& I \to S, \text{recover, rate constant} = c_2^{(d)},\\
& S \to I, \text{infection from outside, rate constant} = c_3^{(d)}.
\end{align*}

To model SIS dynamics at the individual level, we mark the infectious and susceptible individuals with person indexes: at any time, a person $p$ is either an infectious individual denoted as $I^{(p)}$ or a susceptible individual denoted as $S^{(p)}$, but not both. We further define $i^{(p)}$ as a binary random variable that is $1$ when person $p$ is infectious and $0$ when the person is susceptible, and similarly define $s^{(p)}$ as another binary random variable that is $1$ when person $p$ is susceptible and $0$ when the person is not susceptible. Thus $i^{(p)}\in\{0,1\}$, $s^{(p)}\in\{0,1\}$ and $i^{(p)} + s^{(p)} = 1$. The probability for a susceptible person $p$ to become infectious through one unit time of contact with an infectious person $q$ is therefore $h(I^{(p)}+S^{(q)}\to I^{(p)}+I^{(q)}; c_1^{(d)}) = c_1^{(d)}\cdot s^{(p)} i^{(q)} = c_1^{(d)}$.

Conditioned on the instantaneous spatial-proximity network, the reported symptoms of the PocketCare users and of their local proximity network neighbors serve as independent observations.

Trip generation has been a well-studied field since H{\"a}gerstrand's constraint-based modeling \cite{hagerstrand1987human} and Chapin's activity-based modeling \cite{chapin1968activity}. In a trip-generation model, people engage in different activities and trips are undertaken to fulfill these activities. These include major activities such as being at home, working, shopping, being at school, eating out, socializing, enjoying recreation, and serving passengers (i.e., picking up and dropping off), as well as numerous other activities that people engage in on a less-than-daily or even weekly basis, such as going to the doctor or the bank. The trips and activities are dependent on many factors and are traditionally simulated from surveyed results. However, with the ability to survey the travels of a large population with mobile phones, and the availability of large databases, a new method of trip generation combining machine learning and signal processing is possible. 

To model the human dynamics in performing a sequence of activities (e.g., taking courses, performing research, staying in an apartment, dining in a cafeteria, etc.) throughout the day, we characterize people with their locations. Let $l_i$, $l_j$, and $p_k$ denote locations $i$ and $j$ and person $k$. The system is driven by a single type of event, $p_{k}\circ l_{i}\to p_{k}\circ l_{j}$ --- person $k$ moving from location $i$ to location $j$ with rate constant $c_{l_{i},l_{j}}^{(t)}$ (number of events per unit time). Here we use "$\circ$" to represent a bond: person $k$ binds to location $i$ before the event and binds to location $j$ after the event. Let latent state $x_t^{(l)}$ be the population at location $l$ at time step $t$. Then, this event changes the population at location $l_{i}$ from $x_{t}^{(l_{i})}$ to $x_{t+1}^{(l_{i})}=x_{t}^{(l_{i})}-1$, and changes the population at location $l_{j}$ from $x_{t}^{(l_{j})}$ to $x_{t}^{(l_{j})}+1$. According to this model, an individual stays at location $i$ for an average duration $1/\sum_{j}c_{l_{i},l_{j}}^{(t)}$ and on exit chooses the next location with a probability proportional to the rate constant $c_{l_{i},l_{j}}^{(t)}/\sum_{j'}c_{l_{i},l_{j'}}^{(t)}$. 

We use the WiFi scanning trajectories from PocketCare users and syslog to construct activity durations at each WiFi access point, and the transition probabilities from one access point to another. A number of researchers have characterized the structure of daily trips on university campuses \cite{guan2017fine,dong2011modeling}.

Then, we use the number of smartphones scanned by PocketCare users as independent noisy observations of the populations at different locations, assuming that both the Bluetooth-discoverable devices and the PocketCare users are uniformly sampled from the system. Let $x_0$ be the total population in the system and $y_0$ the total population with Bluetooth-discoverable phones. The probability of observing $y_{t}^{(l_{j})}$ population at location $j$ conditioned on there being $x_{t}^{(l_{j})}$ population in total is
$p(y_{t}^{(l_{j})}|x_{t}^{(l_{j})})=\footnotesize\left(\begin{matrix}x_{t}^{(l_{j})}\\
y_{t}^{(l_{j})}
\end{matrix}\right)\left(\begin{matrix}x_0-x_{t}^{(l_{j})}\\
y_0-y_{t}^{(l_{j})}
\end{matrix}\right)\bigg/\left(\begin{matrix}x_0\\
y_0
\end{matrix}\right)$.
When the total population in the system is large, the percentage of Bluetooth-discoverable population at a given location is roughly the percentage in the system. 

To summarize, we capture the latent social network dynamics with three social diffusion events characterized by rate constants $c_1^{(d)}$, $c_2^{(d)}$, and $c_3^{(d)}$, and activity events characterized by rate constants $c^{(t)}_{l_i,l_j}$. We use volunteers' Bluetooth-scanning of the local proximity network and their surveyed symptom reports as noisy observations to infer the latent dynamics. These two types of rate constant are treated the same, which demonstrates the versatility of the discrete event model in capturing diverse dynamics in a unified framework.

\subsection{Inference and learning}

Although the stochastic kinetic model is a continuous time model, we work with a discrete time stochastic model in the rest of this paper because our goal is to track stochastic kinetic dynamics from observations of populations or individuals with countably many computational steps. To this end, we approximate the continuous time process with a discrete time process on a countable set of equally spaced time points $0,\tau,2\tau,\dots$, with a time interval so small that the probability of more than one event happening in the interval $\tau$ is negligible. This approximation works because the state transition kernel from time $0$ to time $\tau$ is 

\begin{equation}
p(x_{0}\to x_{\tau})=\sum\limits_{n=0}^{\infty}(I+\frac{Q}{\gamma})^{n}\exp(-\gamma\tau)\frac{(\gamma\tau)^{n}}{n!}.\label{eq:pf-uniformization}
\end{equation}

according to the uniformization method \cite{grassman77uniformization}, where $\gamma$ is a uniformization rate, $I$ the identity matrix and $Q$ the infinitesimal generator defined by $h_{k},k=1,\dots,V$. With $\gamma\to\infty$ and $\gamma\tau=1$, we get a first-order approximation of the state transition kernel $I+Q\cdot\tau$.

Specifically, let $v_{1},\dots,v_{T}$ be a sequence of events in the discrete time stochastic kinetic system, $x_{1},\dots,x_{T}$ a sequence of states (populations of species), and $y_{1},\dots,y_{T}$ a set of observations about the populations. Our goal is to make inferences about $\{v_{t},x_{t}:t=1,\dots T\}$ from $\{y_{t}:t=1,\dots,T\}$ according to the following probability measure, where indicator function $\delta(x_{t}-x_{t-1}=\Delta_{v_{t}})$ is 1 if the previous state is $x_{t-1}$ and the current state is $x_{t}=x_{t-1}+\Delta_{v_{t}}$, and 0 otherwise.
\begin{align}
 & P\left(v_{1,\dots,T},x_{1,\dots,T},y_{1,\dots,T}\right)=\prod_{t=1}^{T}P(x_{t},y_{t},v_{t}|x_{t-1}),\label{eq:DTSKM}\\
 & \mbox{where }P(x_{t},y_{t},v_{t}|x_{t-1}) = P(v_{t}|x_{t-1})\delta(x_{t}-x_{t-1}=\Delta_{v_{t}})P(y_{t}|x_{t}),\label{eq:DTSKMTransition}\\
 & \mbox{and }\hspace*{1em}P(v_{t}|x_{t-1})=\begin{cases}
c_{k}\tau\cdot g_{k}\left(x_{t-1}\right) & \mbox{if }v_{t}=k\\
1-\sum_{j}c_{j}\tau g_{j}\left(x_{t-1}\right) & \mbox{if }v_{t}=\emptyset
\end{cases}.\label{eq:DTSKMP}
\end{align}

\begin{figure}[t!]
\center
\includegraphics[width=0.8\columnwidth]{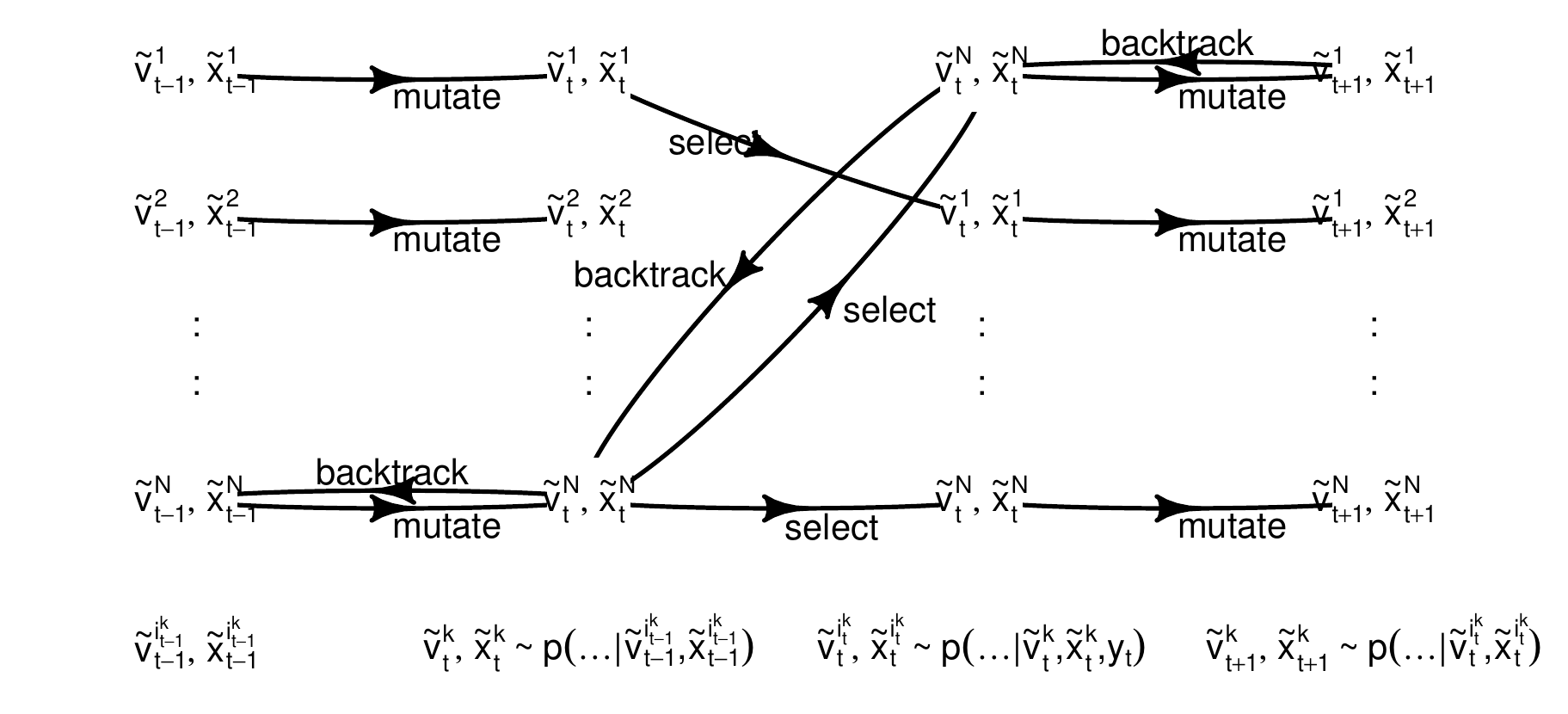}
\caption{A graphical illustration of particle filtering and particle smoothing algorithms.}
\label{fig:pfpm}
\end{figure}

We apply a particle filter to track the dynamics of this process (Figure \ref{fig:pfpm}). The particle filter maintains a collection of particles $x_{t}^{k}$ for $k=1,\cdots,N$ and $t=1,\cdots,T$ to represent the likelihood of the latent state of a stochastic process $x_{t}$ at different regions of the state space with each particle representing a system state, given noisy and partial observations $y_{1,\cdots,T}$. It tracks the evolution of a stochastic process by alternately mutating/sampling the collection of particles according to stochastic process dynamics $P(x_{t}|x_{t-1})$ and selecting/resampling the particles according to observations $P(y_{t}|x_{t})$. Here the particle mutation and selection metaphors are from statistical physics \cite{del2004feynman}, and refer to particle sampling and resampling respectively. In comparison with a particle filter, a simulation run uses only one particle and does not use observations to perform particle selection.

Specifically, let $x_{t}^{k}$ for $k=1,\cdots,N$ be the collection of particle positions and $v_t^k$ the corresponding events from particle mutation, and $i_{t}^{k}\in\{1,\cdots,N\}$ be the collection of particle indexes from particle selection. To make inferences about the latent state $x_{t}$ of a stochastic process starting at state $x_{0}$ from observations $y_{1:t}$, we initialize particle positions and indexes as $x_{0}^{1},\cdots,x_{0}^{N}=x_{0}$ and $i_{0}^{1}=1,\cdots,i_{0}^{N}=N$, and iteratively sample the next event $v_{t}^{k}$ according to how likely it is that different events will occur conditioned on system state $x_{t-1}^{i_{t-1}^{k}}$ for $k=1,\cdots,N$ (Eq. \ref{eq:pf-v}), then update $x_{t}^{k}=x_{t-1}^{k}+\Delta_{v_{t}^{k}}$ accordingly (Eq. \ref{eq:pf-x}) and resample these events per their likelihoods with regard to the observation $y_{t}$ for $t=1,\cdots,T$ (Eq. \ref{eq:pf-i}).
\begin{align}
 & v_{t}^{k}|x_{t-1}^{i_{t-1}^{k}}\sim\text{Categorical}(1-\frac{h_{0}(x_{t-1}^{i_{t-1}^{k}})}{\gamma},\frac{h_{1}(x_{t-1}^{i_{t-1}^{k}})}{\gamma},\cdots,\frac{h_{V}(x_{t-1}^{i_{t-1}^{k}})}{\gamma}),\label{eq:pf-v}\\
 & x_{t}^{k}=x_{t-1}^{i_{t-1}^{k}}+\Delta_{v_{t}^{k}},\label{eq:pf-x}\\
 & i_{t}^{k}|(x_{t}^{1:N},y_{t})\sim\text{Categorical}(p(y_{t}|x_{t}^{1}),\cdots,p(y_{t}|x_{t}^{N})).\label{eq:pf-i}
\end{align}

To determine a particle trajectory from the posterior distribution of a stochastic kinetic process with respect to observations, we trace back the events that led to the particles $x_{T}^{i_{T}^{k}}$ for $k=1,\cdots,N$: 
\begin{align}
 & x_{0},v_{1}^{j_{1}^{k}},x_{1}^{j_{1}^{k}},\cdots,v_{T}^{j_{T}^{k}},x_{T}^{j_{T}^{k}},\text{where }{\scriptstyle j_{T}^{k}=i_{T}^{k},j_{T-1}^{k}=i_{T-1}^{j_{T}^{k}},\cdots,j_{1}^{k}=i_{1}^{j_{2}^{k}}}.\label{eq:pf-path} 
\end{align}
To learn the rate constants (the parameters) of a stochastic kinetic model, we sample from the posterior distribution of these parameters conditioned on the particle trajectory, using a beta distribution as conjugate prior. Let $v_{1},\cdots,v_{T}$ be the sequence of events in a particle trajectory (Eq. \ref{eq:pf-path}). The posterior probability distribution of a rate constant is a beta distribution that matches the expected number of events in a sample path with the number of events that actually occur, where $a_v$ and $b_v$ are hyper-parameters: 
\begin{align}
&c_{v}|v_{1:T},x_{1:T}\sim\text{Beta}(a_{v}+\sum\nolimits_{i=1}^{T}\delta_{v}(v_{t}),b_{v}+\sum\nolimits_{t=0}^{T}g_{v}(x_{t})-\sum\nolimits_{t=1}^{T}\delta_{v}(v_{t}) ).\label{eq:pf-learn}\\
\end{align}

Overall, we have developed a particle-based algorithm to make inferences about a complex system using a stochastic kinetic model and noisy observations (Algorithm \ref{alg:PF}). This algorithm gives social scientists a way to track real-world complex social systems from continued noisy observations with their simulator, as long as they can evolve the system state with the simulator and compute the likelihood of the system state with respect to observations. When the complex system has an extremely high dimension, variational inference algorithms \cite{xu2016using,fang2017expectation, yang2019optimal} should be considered to avoid particle degeneracy issues. However, as our experiment shows, the particle-based algorithm works well with social systems under general considerations.

\begin{algorithm}[t]

\textbf{Input}: Observations $y_{1},\cdots,y_{T}$
of a stochastic kinetic process (Eq. \ref{eq:DTSKM}) specified by
a set of events (Eq. \ref{eq:reaction} and \ref{eq:DTSKMP} for $v=1,\cdots,V$).

\textbf{Output}: Resampled particles $(v_{t}^{i_{t}^{k}},x_{t}^{i_{t}^{k}})_{t=1:T}^{k=1:N}$
from particle filter, particle trajectories $(v_{t}^{j_{t}^{k}},x_{t}^{j_{t}^{k}})_{t=1:T}^{k=1:N}$
from particle smoother.
\vspace*{.5em}

\textbf{Procedure}: 

\begin{itemize}
\item Initialize $x_{0}^{1}=\cdots=x_{0}^{N}=x_{0}$, $i_{0}^{1}=1,\cdots,i_{0}^{N}=N$.
\item (Filtering) For $t$ in $1,\cdots,n$ and $k$ in $1,\cdots,N$, sample $v_{t}^{k}$
and $i_{t}^{k}$ according to Eq. \ref{eq:pf-v}, \ref{eq:pf-x} and
\ref{eq:pf-i}, where $p(y_{t}|x_{t})$ is defined in Eq. \ref{eq:DTSKM}.
\item (Smoothing) Back-track particle trajectory from $x_{T}^{i_{T}^{k}}$
according to Eq. \ref{eq:pf-path}, for $k=1,\cdots,N$.
\item (Parameter Learning) Sample rate constants according to Eq. \ref{eq:pf-learn}.
\end{itemize}
\caption{\label{alg:PF}Particle-Based Inference with Stochastic Kinetic Model}
\end{algorithm}

\section{Results}

In this section, we establish the statistical significance of observing symptom propagation in a physical-proximity network, then we evaluate the performance of predicting infection at both the individual level and the population level.

\subsection{Feasibility of tracking a proximity network with a small number of volunteers}

In this data set, interactions happens at places of with high population densityies. For example, students take classes at lecture halls, research teams meet in department buildings, and people take meals together at the commons and or do work outs at the gym. Hence, Iit is possible to track the activities and interactions of a large population with a small number of volunteers using their mobile phones to Bluetooth-scan nearby mobile phones because p. People performed similar activities daily at their regular times and locations,. which allowsHence it is possible to modeling the activities and interactions of non-volunteers from those of the volunteers of the same types (e.g., students and the facultyies of different departments). 

\begin{figure}[]
   \begin{minipage}{0.49\textwidth}
     \centering
     \includegraphics[width=1\linewidth]{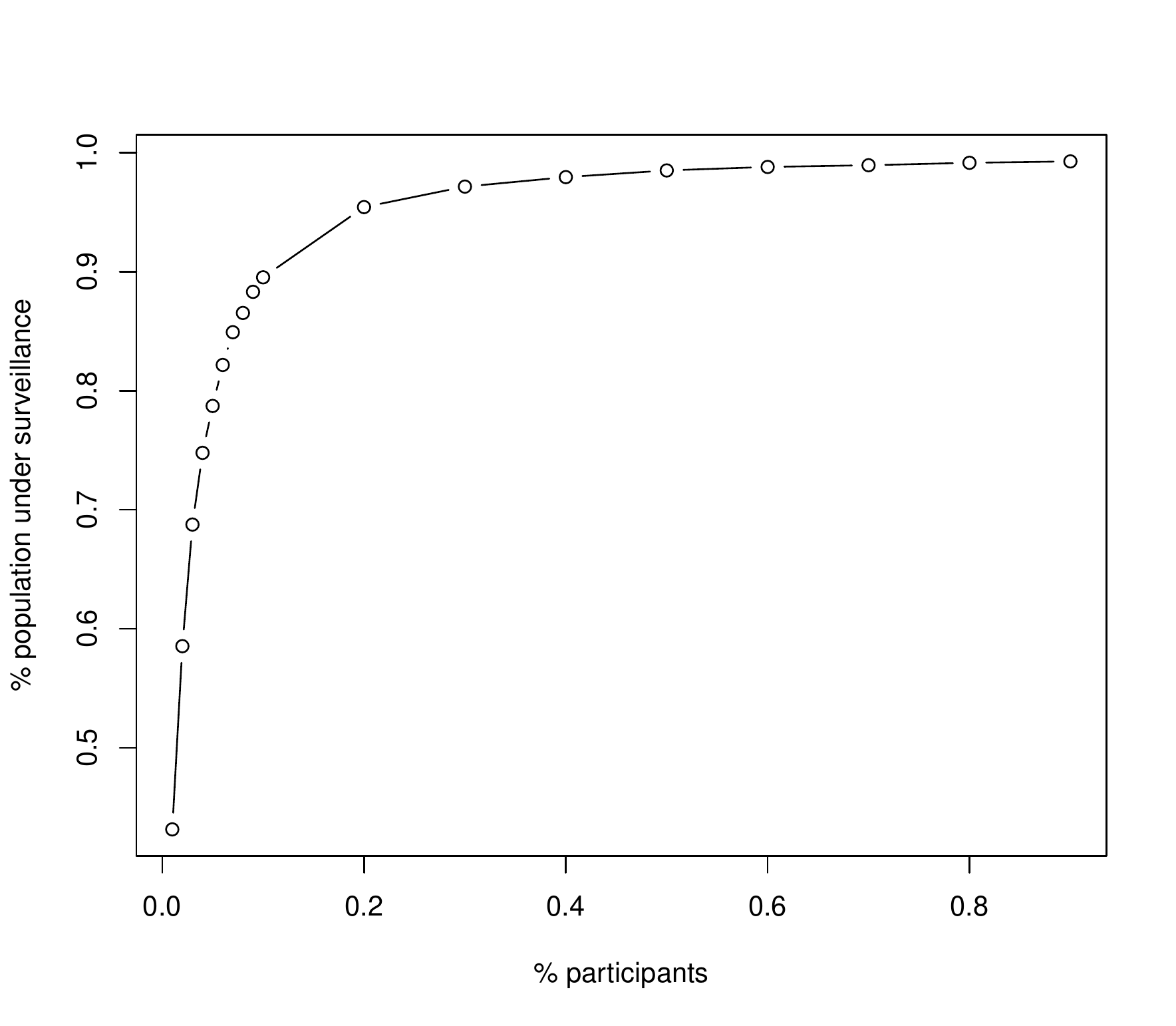}
     \caption{Observing dynamic proximity-network with\\ a small number of volunteers}\label{Fig:survey-community-network}
   \end{minipage}\hfill
   \begin{minipage}{0.49\textwidth}
     \centering
     \includegraphics[width=1\linewidth]{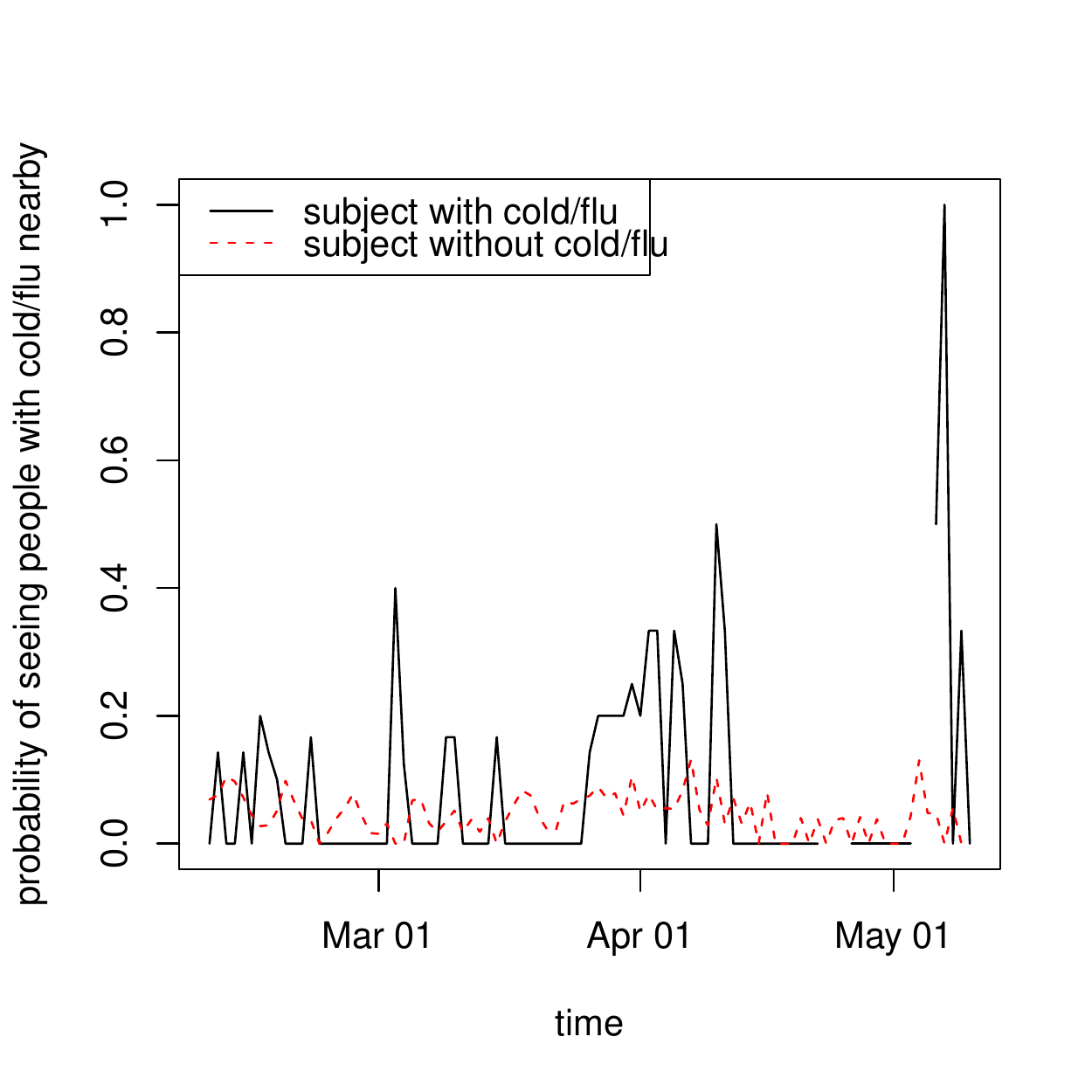}
     \caption{The probabilities for infectious and healthy individuals to report nearby infectious cases}\label{Fig:infection:net_effect}
   \end{minipage}\hfill
\end{figure}

In Figure \ref{Fig:survey-community-network}, we show that we can randomly sample 5\% of the population as observers to detect the presence of up to 80\% of the population who use the same WiFi access points at the same time on a university campus of around 30 30,000thousand people. If we randomly sample 10\% of the population as observers, the presence and activities of 90\% of the population can be observed. The data to generate this figure is are the syslog from the university-owned Wi-Fi access points containing which anonymized device IDs are connected to which Wi-Fi access points for what times and for how long.

\subsection{Statistical significance of symptom propagation}

In our data, a PocketCare user with a symptom has 2 two times higher odds  of seeing other users in his spatial-proximity network with the same symptom (Figure \ref{Fig:infection:net_effect}). As such, it makes sense to fit the time-tested infection model with real-world data of symptom reports and proximity observations, and infer how people infect one another through their contacts. In order to determine whether the higher odds could somehow be due to chance, we have conducted the following permutation test to reject the null hypothesis that "the spatial-proximity network is unrelated to symptoms,", and we can reject that null hypothesis with $p<0.01$. The permutation test shuffles the mapping between the users and the nodes in the proximity network and estimates the probability distribution of the number of neighbors with the same symptom among all possible shuffling outcomes. If proximity networks are not related to the timing of when a user exhibits a symptom, then all mappings between the users and the nodes would be equally likely, and the number of friends with the same symptom would take the more likely values.

Further, the probability of exhibiting a symptom increases approximately linearly with the number of spatial -contacts reporting the same symptom. The base probability of having a symptom is 4 cases per day per 100 persons, and every additional friend exhibiting the symptom adds increases the probability by about 1\%. This relationship again agrees with the theory of epidemic dynamics, which predicts that the rate of contagion will be proportional to the likelihood of contact with an infected individual.

The epidemic model says that an infected/infectious individual recovers with a constant rate $\gamma$, and therefore the duration of the infection follows an exponential distribution. We fit the observed symptom durations with an exponential distribution using maximum likelihood estimation to check the compatibility of the epidemic model with the collected data on recovery rate. The average duration of the symptoms in the fitted exponential distributions is about 3 three days. As such, we have satisfactory goodness-of-fit statistics in Kolmogorov-Smirnov hypothesis testing.

In Figure \ref{Fig:epidemic-progression}, we show the stochastic matrix of common cold and flu progression generated from the surveyed daily symptoms reports about from the volunteers, with the states ordered according to their times of occurrence in a disease episode. The progression of a common cold/flu episode starts from with runny nose, advances to coughing, soare throat, and fever, and ends with normal state. In the experiment, we have asked the subjects to report daily whether they have fever, runny nose, coughing, sore throat and nausea, and we code their responses as a 5-bit string, with 1 bit per symptom. For example, if a subject reports runny nose and coughing, we code the response as "01100.". To generate the stochastic matrix, we tabulate the responses of the volunteers on the current day and the next day, and from the table we estimate the probability distributions of the response in the next day conditioned on the response of the current day. To order the responses according to their relative positions in a common cold/flu episode, we simulate common cold/flu progression in an episode from the stochastic matrix and estimate the average times for the responses to occur. Thus, each row of the stochastic matrix is identified by the response of the current day, and the entries in the row are the probabilities of the responses in the next day.\\

\subsection{Configuration for evaluating epidemic-tracking performance}
Using the noisy observations of local proximity networks through volunteers' Bluetooth scanning of nearby mobile phones and the reported symptoms about and around the volunteers, we are interested in the following predictions: 1) the probability for individuals to be infected in the near future, given their positions relative to the volunteers in the spatial-proximity network, 2) the smoothing probability distribution of a sample path given the observations, and 3) the total number of infectious people in the near future. Prediction (1) is important for healthcare policy researchers in order to optimally allocate limited resources for controlling epidemic spreading. Prediction (2) is important for calibrating model parameters, while (3) is important for researchers to predict outbreaks ahead of time. 

We use 10-fold cross-validation to evaluate the contributions of the observed spatial-proximity network and of the symptom reporting in predicting whether an individual is infectious two weeks ahead of time, and in inferring whether an individual is infectious on a given day using the available data before and after that day. In other words, we split the 300 volunteers into 10 equal partitions and report the average performance, making predictions on a single partition of held-out volunteers using the other 9 partitions of volunteers. The ground truth is whether the participants reported sick from their daily health surveys. To evaluate the contribution of symptom reports by the volunteers, we remove the daily health survey records (self-sick, nearby-sick, and symptoms) of the held-out volunteers to be predicted, use these as ground truth, and keep the spatial-proximity networks of these volunteers in the prediction. To evaluate the contribution of symptoms as well as the proximity-network structure reported by the volunteers, we take out not only their daily health survey records but also their spatial-proximity networks. So, in the latter case the proximity information about a held-out volunteer is only as complete as what can be observed by the other 9-fold volunteers. 

We draw the precision-recall curve in different configurations, and report the average precision as a performance metric. {\color{red}A precision-recall curve shows the capability of a binary classifier to detect the positive cases as the discrimination threshold of this classifier is varied.} Here, precision is the probability for an individual to actually be infectious (positive) when we predict infectiousness, recall is the fraction of the actually infectious individuals we can identify in the system, and the usage of different thresholds gives different precision-recall pairs. In a precision-recall plot, the {\color{red}no-skill (random-guess)} line is the total number of positive cases divided by the total number of positive (infectious) and negative (non-infectious) cases, which is precision = 0.04. Perfect skill corresponds with precision = 1 and recall = 1. {\color{red}The average precision summarizes the skill of the binary classifier over the whole precision-recall curve, and is computed as the area under the precision-recall curve when recall spans from zero to one.} A precision-recall curve is preferred to a receiver operating characteristic curve because we are less interested in predicting the negative cases correctly when there is an overwhelmingly large fraction of them. {\color{red}Another way to show the ability of a binary classifier is to plot error (i.e. $1 - \text{precision}$) versus recall, resulting a curve that is visually similar to a receiver operating characteristic (ROC) curve. We will use the more well-known precision-recall curve in the following. The precision-recall curve is a stochastic process with probability dependence on both the discriminant function of the binary classifier, and the training/testing data sampled from an underlying probability distribution. The curves could take rugged shapes when they are drawn from insufficient amount of data. Depending on many factors, sometimes they can cross one another.}

We use bootstrapping to evaluate the performance of predicting daily infectious-population size two weeks ahead of time, because it is not feasible to mobilize an entire population of several thousands to establish the ground truth. To this end, we construct a sample path within the dynamic proximity network of the whole community and epidemic diffusion from the partial observations using the particle smoothing algorithm. More specifically, we sample 10\% of the population as volunteers, and then replay the sample path to predict the infectious populations in a two-week time window using only the partial observations up to the current time at each time point. Then we repeat this procedure, obtaining many synthesized ground-truth sample paths and estimated daily infectious populations from partial observations, from which we get the sample averages.

By means of the Monte Carlo integration just described, we identify the contribution of the partially observed proximity network in predicting the total infectious population two weeks ahead of time by comparing our discrete event model particle filter with a scaling-based algorithm and a support vector regression algorithm, where the particle filter algorithm uses the information contained in the partially observed proximity network and the latter two baseline approaches do not. Instead, the scaling-based algorithm uses the percentage of the respondents reporting symptoms as a surrogate for the community-wide percentage of infection, as is commonly used by healthcare researchers. The support vector regression algorithm predicts the community-wide percentage of infection in a two-week time window from the percentage of respondents reporting symptoms about themselves and the people around them in the previous seven days. On the other side, the discrete event model particle filter algorithm simulates network and social diffusion dynamics at the individual level in agreement with the reported symptoms and the local proximity-network structure. 

In the following section, we discuss the r-squared statistics and visually compare the predicted infectious populations from noisy observations using these three algorithms (the discrete event model, a scaling-based algorithm, and support vector regression) with the ground truth in the same plot. Let $f_{t}$ be the predicted infectious population at time $t$ calculated from the information until two weeks before time $t$, $y_{t}$ the ground truth, and $\bar{y}$ the average of $y_{t}$. We define $R^{2}=1-\sum_{t}(f_{t}-y_{t})^{2}/\sum_{t}(y_{t}-\bar{y}_{t})^{2}$. A higher $R^{2}$ indicates a better fit between the estimated time series and the ground truth, with $R^{2}=1$ indicating a perfect fit and $R^{2}<0$ a fit worse than using the average.

We also use a stress test to calibrate the minimum requirement to predict infection at the individual and system levels from volunteer-reported local spatial-proximity networks and symptoms of infection. To this end, we randomly select 150, 100, and 30 volunteers from the set of 300 and evaluate prediction performance at the individual and aggregate levels. Asymptotically, removing volunteers is equivalent to having volunteers not contributing data.\\

\subsection{Performance in tracking an epidemic}
In this section, we report the performance in predicting infection at both the individual level and the aggregate level from the signs of infection and the local spatial-proximity network structure observed by the volunteers.

\begin{figure}[]
   \begin{minipage}{0.49\textwidth}
     \centering
     \includegraphics[width=1\linewidth]{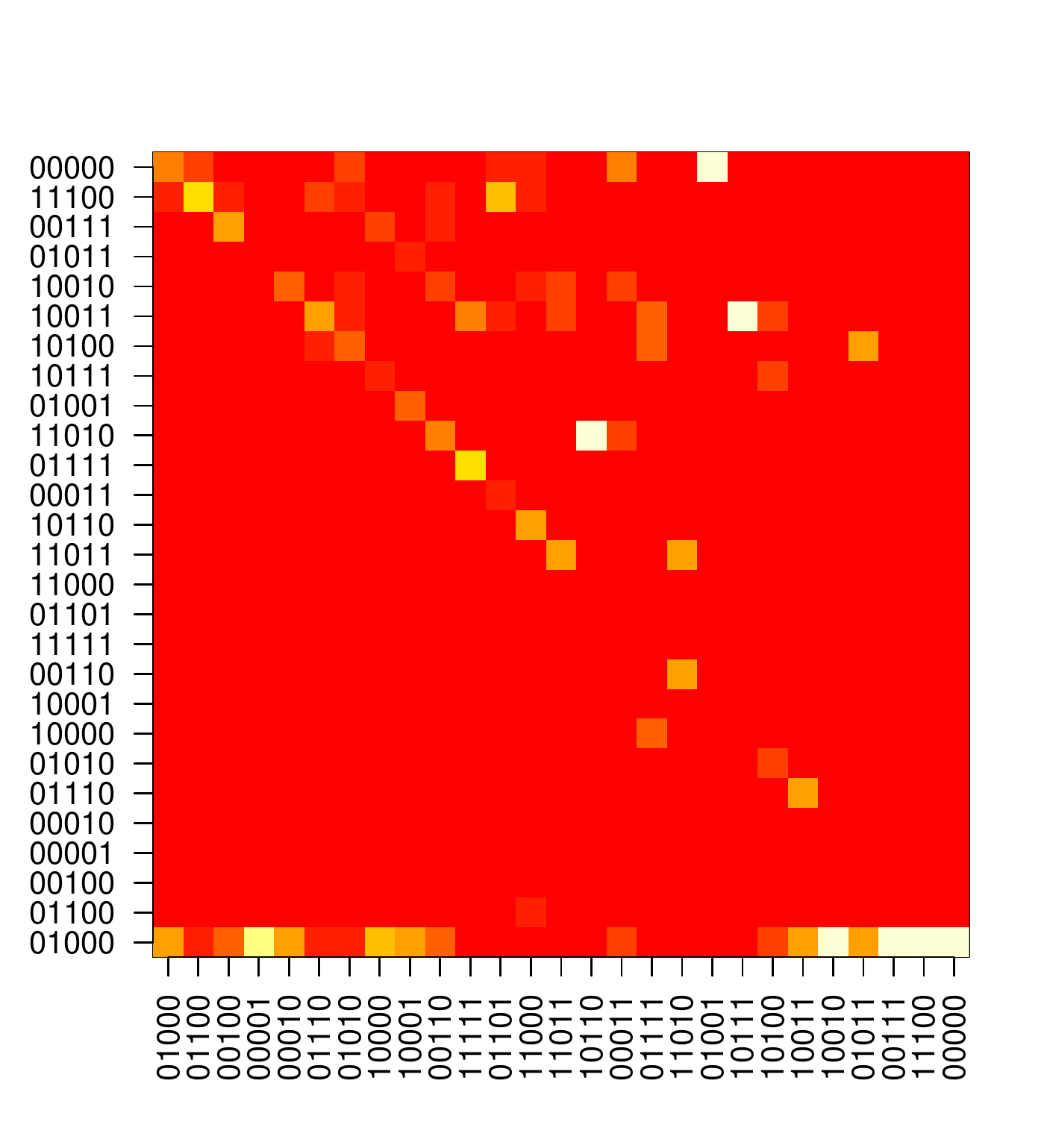}
     \caption{Stochastic matrix of symptom development\\ constructed from survey data}\label{Fig:epidemic-progression}
   \end{minipage}
   \begin{minipage}{0.49\textwidth}
      \centering
	  \includegraphics[width=1\textwidth]{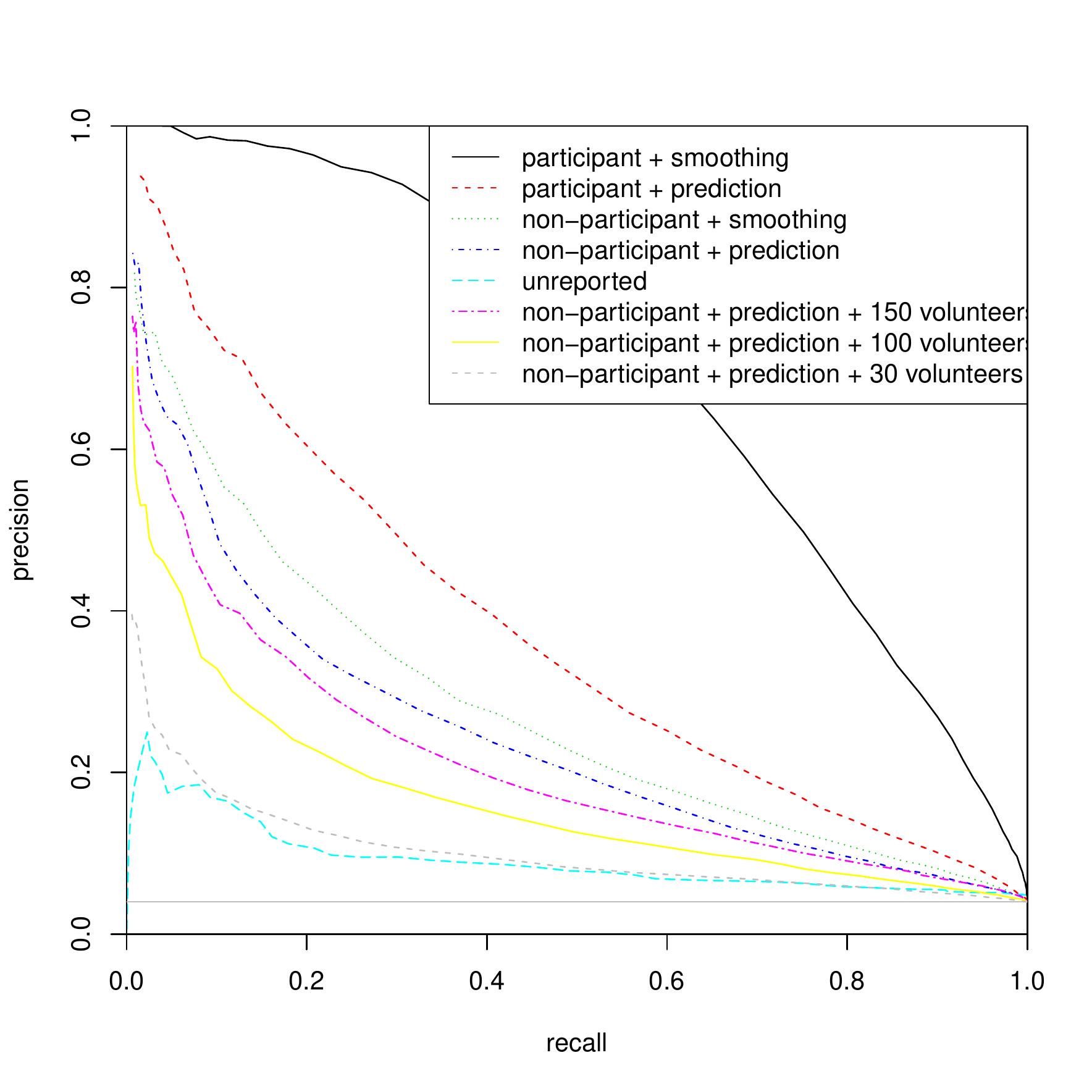}
      \caption{\label{fig:performance-roc}Precision-recall curves of predicting infection in different configurations}
  \end{minipage}\hfill
   \begin{minipage}{0.40\textwidth}
      \centering
	  \includegraphics[width=1\textwidth]{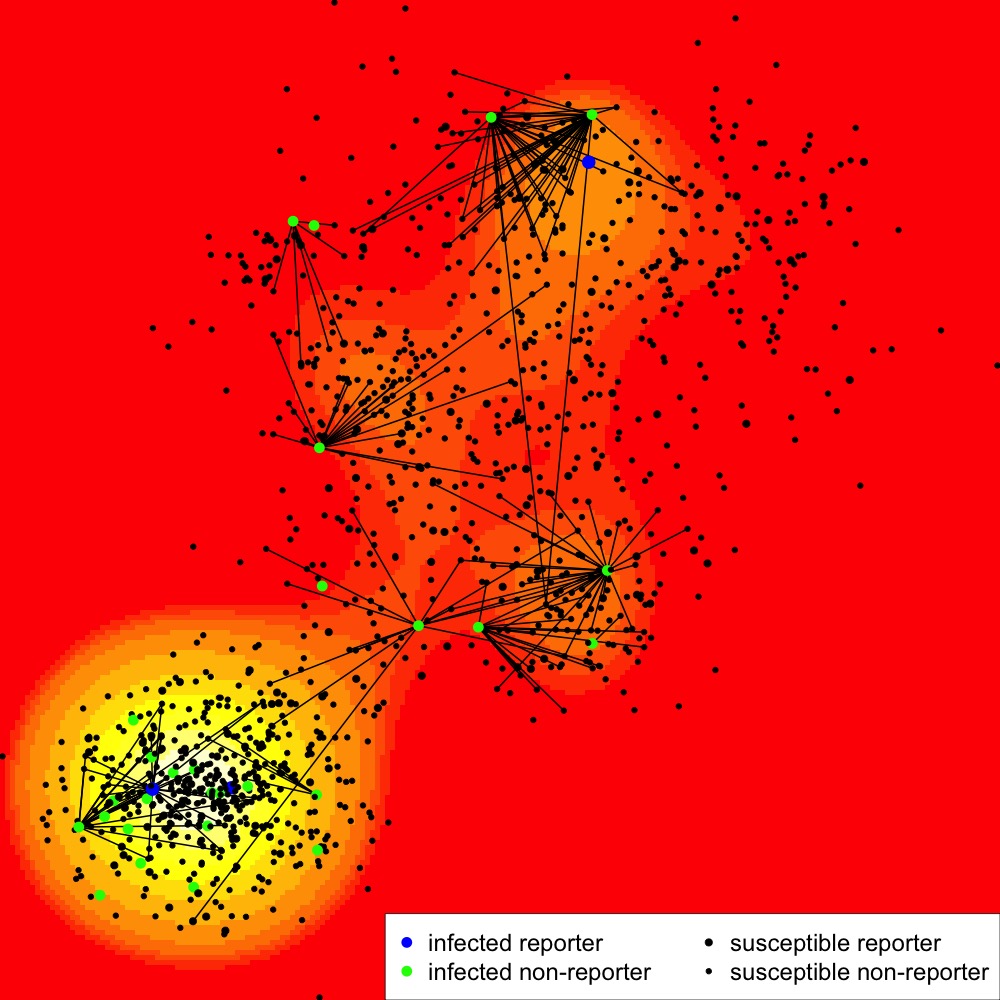}
      \caption{\label{fig:inference-snapshot}Snapshot of particle-smoothing inference}
  \end{minipage}\hfill
  \begin{minipage}{0.49\textwidth}
      \centering
      \includegraphics[width=1\textwidth]{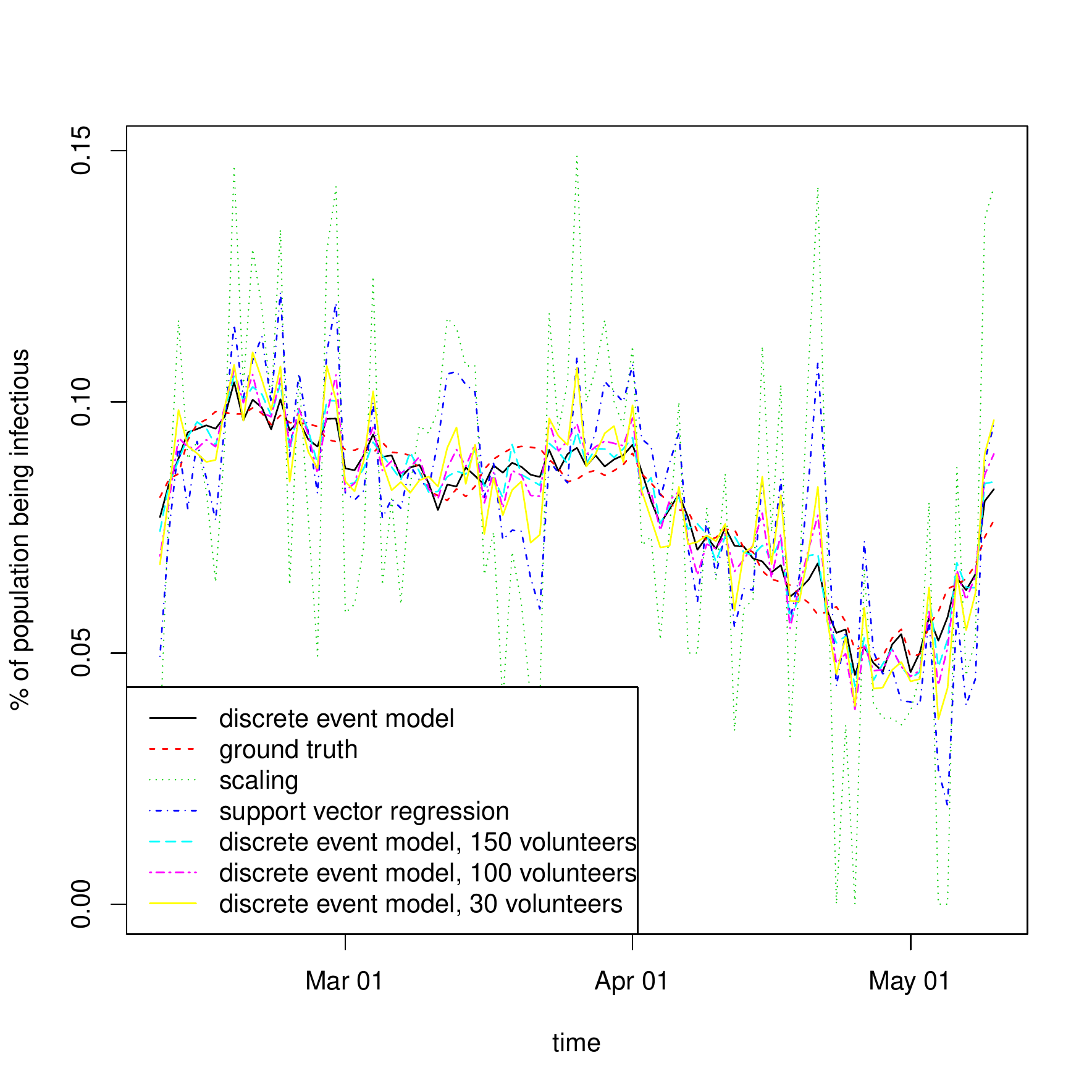}
      \caption{\label{fig:performance-aggregate}Predicting percentage of population being infectious in 2-week window}
  \end{minipage}

\end{figure}

Figure \ref{fig:performance-roc} shows in different configurations the precision-recall curves of predicting whether an individual is infectious from the reported symptoms and the local proximity network contributed by PocketCare users. The goal of this comparison is to establish the contributions of the various pieces of observations in predicting whether an individual is infectious. In this plot, the \emph{prediction} curves correspond to prediction of probability of infection two weeks ahead of time, the \emph{smoothing} curves correspond to estimating whether an individual is infectious based on observations from other individuals, and the \emph{unreported} curve corresponds to the prediction from only the reported symptoms without using the proximity network. The \emph{participant} curves correspond to the prediction/smoothing of infectious status using the subject's own local proximity network through Bluetooth scanning, and the \emph{non-participant} curves correspond to the computation using the proximity networks of other people. The curves marked with 150, 100, and 30 volunteers correspond to the usage of observations from 5\%, 3.3\%, and 1\% of the population to make predictions. 

The {\color{red}random-guess} curve corresponds to precision = 0.04, and we can infer whether a volunteer reported being sick in the health survey with an average precision = 0.70 given all observations before and after the day of prediction (\emph{participant + smoothing}). Given only the fraction of volunteers who reported symptoms, we can achieve an average precision of 0.09 in predicting infection in a two-week time window (\emph{unreported}), which is twice the {\color{red}random-guess} baseline. Given the most complete information about an individual's local spatial-proximity network --- which is the situation where the individual contributes to Bluetooth-scanning the data but not to symptom surveys --- we can achieve an average precision of 0.35 (\emph{participant + prediction}), which is nine times the baseline. With incomplete proximity-network information observed by volunteers, we can achieve an average precision of 0.24 (\emph{non-participant + prediction}), which is six times the baseline, and we can predict the 10\% most likely infection cases (where recall $\le$ 0.1) with 60\% precision. When we drop the fraction of volunteers who report spatial-proximity networks and symptoms of infection to 5\%, 3.3\%, and 1\%, the average precision first drops slightly to 0.22, then collapses to 0.17 and 0.10. The observations about the proximity network improve the performance of individual-level prediction significantly, because infectious individuals make different contributions to epidemic progression. Indeed, infectious diseases from randomly infected individuals go first to the hubs of a social network then spread to the other nodes starting from those hubs.

Figure \ref{fig:inference-snapshot} shows a snapshot of particle-smoothing inference from partial observations of volunteer symptoms and proximity-networks. In this figure, we use points of bigger and smaller sizes to represent the volunteers and the non-volunteers in the community, and embed the proximity network in a two-dimensional space according to the distance needed for a disease to travel from one person to another person. We use a heat-map in the background to represent the likelihood for a person in the area to be infectious --- that is, there are bigger fractions of infectious people in "hotter" regions. 

While we do not have the complete proximity network in the inference of epidemic diffusion, we can nevertheless construct structurally equivalent networks concerning epidemic diffusion dynamics. Similarly, in inferring the likelihood for a volunteer or non-volunteer to be infectious, all we need is that individual's relative positions with respect to other volunteers in the proximity network. In other words, two people sharing the same position in the network will have the same likelihood of being infectious. As a result, the PocketCare framework can predict infection for both the volunteers and the non-volunteers, as long as they can specify to what extent they interact with the volunteers.

Finally, Figure \ref{fig:performance-aggregate} compares the performance of predicting the infectious population in a two-week time window every day, starting from recruiting 10\% of the population as volunteers to report daily partial observations about the spatial-proximity network and epidemic diffusion. In other words, each prediction is based on observations up to two weeks before. The partial observation of the dynamic proximity network is made through Bluetooth-scanning of nearby mobile phones, and the partial observation of symptom diffusion through the symptoms of the volunteers themselves and of the other people spotted by the volunteers. The discrete event model particle filter has an r-squared value of 95\%, in comparison with 60\% for support vector regression and 20\% for a scaling-based method. When we drop the volunteer population to 150, 100, or 30 (corresponding to 5\%, 3.3\%, and 1\% of the total population), we can still achieve an r-squared value of 90\%, 80\%, or 70\%, respectively. So, even 1\% of the population can capture the structure of the spatial-proximity network and the dynamics of epidemic spreading reasonably well at the aggregate level. Conversely, a scaling-based method has a large deviation from the ground truth because the respondent population is small in comparison with the population to monitor. A model-free approach such as support vector regression still has a large deviation from the ground truth. This is because not all infectious individuals contribute to epidemic progression in the same way.

\section{Discussion and Conclusions}
In the current paper, we have described a multi-disciplinary approach of monitoring the spatial-proximity interactions among individuals in a community by mobilizing a small fraction of the community as volunteers to detect other people with mobile-phone Bluetooth-scanning and to report signs of epidemic diffusion with mobile-phone surveys. To stitch together the isolated observations of the volunteers and the latent interactions involving the non-volunteers with high-fidelity social interaction models, we derived inference and learning algorithms for a discrete event model, based on the premise that many complex social-interaction dynamics can be factored into a sequence of elementary events that individually involve only a few people but together describe the complex behavior of the system. We have applied our approach to the prediction of common cold and flu propagation in a community of 3000 people using the noisy observations of spatial-proximity interactions from 300 volunteers randomly selected from the community. The results show that we can predict common cold and flu infection two weeks ahead of time with an average precision from 0.24 to 0.35 depending on the amount of information. This is six to nine times the precision that a  {\color{red}random-guess} model can achieve. Moreover, we can predict infectious population in a two-weeks window with an r-squared value of 0.95, in comparison with an r-squared value of 0.2 for a {\color{red}random-guess} model.

A fundamental challenge in applying ubiquitous-computing technologies to sense social dynamics is estimating diverse statistics from heterogeneous data sources, including for example 1) location observations of mobile phone users, 2) various dynamic networks defined by spatial proximity, phone calls, and SMSs, and 3) attribute assessments about individuals. Usually, estimation is achieved either by training a machine learning model to map hand-crafted features to the desired statistics or by running a simulator calibrated with data to generate trajectories for calculating statistics. Is there a principled method for estimating these statistics from a common set of data, hypotheses, or machine learning algorithms? In this paper, we have proposed an approach that infers the posterior distribution of complex system dynamics with high resolution and fidelity by connecting the dots of isolated observations using simulation models of social dynamics and machine learning, and applies the inferred individual-level dynamics to calculate the desired variables from cross-domain observations and dynamics and to identify actionable insights and solutions. The proposed approach is desirable because it provides an affordable microscope with which researchers can conduct accountable and repeatable experiments in living-lab settings.

Our multi-disciplinary approach is based on several working hypotheses. Theoretically, we hypothesized that complex system dynamics could be factorized into a sequence of known elementary events with each involving the interaction of only a few individuals. This hypothesis is strongly supported by our results and by the wide application of various modeling languages in capturing discrete event dynamics, including the stochastic Petri-net, stock and flow diagram, continuous time probabilistic model, and compound Poisson process. In an uncharted domain where the discrete event dynamics can be unknown, automatically learning a sparse discrete event representation of the unknown dynamics poses an interesting problem and parallels the trend of machine learning research, where novel sources of data demand more powerful algorithms to learn representations that approach the expressiveness of a simulation model. Practically, we hypothesized that a significant fraction of the community setting their mobile phones to be Bluetooth-discoverable and their Wi-Fi to be constantly scanning nearby access points. This hypothesis is also strongly supported by our results discussed. 

\begin{acks}
{\color{red}We thank the anonymous reviewers for improving the manuscript. Research reported here was supported by a University at Buffalo Innovative Micro-Programs Accelerating Collaboration in Themes (IMPACT) award.}
\end{acks}

\bibliographystyle{ACM-Reference-Format}
\bibliography{sample}

\end{document}